\documentclass[journal]{vgtc}                     

\onlineid{0}
\vgtccategory{Research}

\title{Understanding How Humans Inject Knowledge into Machine Learning Workflows through Visual Analytics}

\author{%
  \authororcid{Yiwen\ Xing}{0000-0003-1521-6616},
  \authororcid{Philip\ Beaucamp}{0009-0007-2106-9516},
  \authororcid{Joyraj\ Chakraborty}{0000-0001-7609-1277},
  \authororcid{Afrah\ Farea}{0000-0003-4412-5377},\\
  \authororcid{Yuanzhe\ Jin}{0000-0002-6910-580X},
  \authororcid{Saiful\ Khan}{0000-0002-6796-5670},
  \authororcid{Gennady\ Andrienko}{0000-0002-8574-6295},
  \authororcid{Natalia\ Andrienko}{0000-0003-3313-1560},
  and 
  \authororcid{Min\ Chen}{0000-0001-5320-5729}
}

\authorfooter{
  \item Y. Xing, P. Beaucamp, J. Chakraborty, Y. Jin, and M. Chen are with University of Oxford, UK. E-mail: \{yiwen.xing, philip.beaucamp, joyraj.chakraborty, yuanzhe.jin, min.chen\}@eng.ox.ac.uk,
  \item A. Farea is with Istanbul Technical University, Turkey. E-mail: \{farea16@itu.edu.tr\},
  \item S. Khan is with the Science and Technology Facilities Council, UK. E-mail: saiful.khan@stfc.ac.uk,
  \item G. Andrienko and N. Andrienko are with Fraunhofer Institute IAIS, Germany and City St George's University of London, UK. E-mail: \{gennady,~natalia\}.andrienko@iais.fraunhofer.de.
}

\abstract{%
Visual analytics (VA) plays an increasingly important role in supporting machine learning (ML) workflows. In the field of visualization, such approaches and techniques are referred to as VIS4ML. While ML models are mostly learned automatically, the corresponding ML workflows receive a variety of human inputs, such as data labelling, feature engineering, model architecture designing, hyper-parameter tuning, and so on. 
In this work, we surveyed over 200 VIS4ML papers to gain an understanding of how humans inject their knowledge into ML workflows through interactive visualization. We collected a corpus of VIS4ML papers from the IEEE VIS conferences in the past decade. We developed a coding scheme to facilitate the literature research from four perspectives: characteristics of ML, visualization, interaction, and actions. The analysis of the coded dataset allows us to observe different pathways that transfer human knowledge to ML workflows via interactive visualization. Building on the analysis, we explain the phenomena of VIS4ML using the conceptual model that views VA as model building and the information-theoretic cost-benefit analysis that reasons VA as for optimizing ML workflows. This work provides unequivocal evidence showing the merits of using VA in ML workflows. \textcolor{black}{The full list of surveyed papers, along with all analysis results and figures, is available at \url{https://vis4ml4hd.github.io/ml-knowledge-inject-va/}}.
}

\keywords{VIS4ML, visual analytics, machine learning, workflow, knowledge.}

\graphicspath{{Figures/}{figs/}{figures/}{pictures/}{images/}{./}} 

\usepackage{tabu}                      
\usepackage{booktabs}                  
\usepackage{lipsum}                    
\usepackage{mwe}                       
\usepackage{ccicons}                   
\usepackage{mathptmx}                  
\usepackage{tabularx}
\usepackage{multirow}
\usepackage{array}

\begin{document}

\firstsection{Introduction}

\maketitle
Machine learning (ML) has become a fundamental technology across a wide range of domains and applications. While modern ML models are increasingly capable of learning complex patterns from large-scale data in an automated manner, there is a growing perception that higher levels of automation alone lead to better models. This view, however, often overlooks the critical role of human knowledge in shaping effective ML systems. In practice, the overall ML workflow remains inherently human-centered. Human expertise is involved at multiple stages, such as data curation, feature engineering, model selection, parameter tuning, and result interpretation. The development of high-quality ML models is therefore not purely an automated process, but one in which human knowledge is continuously injected into the workflow, often in subtle and iterative ways. 

Visual analytics (VA) has been widely adopted as a powerful paradigm for supporting human involvement in ML workflows. By combining interactive visualization with analytical computation, VA systems enable users to explore data, understand model behavior, and refine ML processes. A growing body of research in the visualization community, often referred to as VIS4ML, has demonstrated the benefits of integrating visualization and interaction into different stages of ML workflows~\cite{Sacha2019TVCG}. This line of research has driven the development of VA systems that enable users to analyze and understand machine learning models, and in some cases, interactively refine them through human feedback.
Despite this progress, a fundamental question remains insufficiently understood: \textit{where and how is human knowledge injected into machine learning workflows through visual analytics?} Existing VIS4ML application research has primarily focused on system design, visualization techniques, and supported tasks, yet lacks a unified understanding of how human interactions translate into concrete changes in ML processes. While prior surveys have examined VIS4ML from perspectives such as techniques~\cite{Yuan2021}, workflows~\cite{Sacha2019TVCG}, or data-centric operations~\cite{Wang2024TVCG, Wang2019TVCG}, our work takes a complementary perspective by focusing on human knowledge injection pathways. Instead of analyzing specific data, model, or tasks involved, we investigate how human knowledge is transferred into ML workflows through visualization and interaction, and how such knowledge propagates across different stages of the ML process, based on empirical evidence from published works.

To this end, we conduct a survey of VIS4ML research, based on a curated corpus of over 200 papers from IEEE VIS conferences over the past decade. We develop a structured categorization scheme to analyze these works from four complementary perspectives: \textbf{machine learning characteristics}, \textbf{visualization}, \textbf{interaction}, and \textbf{actions}, to systematically capture how VA support human involvement, and how such involvement leads to modifications in ML workflows (Section~\ref{sec:coding_scheme}).

Based on the coded dataset, we perform a detailed analysis to uncover patterns across VIS4ML systems (Section~\ref{sec:coding_results}). In particular, we identify a set of knowledge injection pathways that characterize how knowledge flows through a loop in VIS4ML systems, where ML-derived data is visualized and interpreted by humans, transformed into knowledge, and subsequently translated into actions that modify different components of ML workflows (Section~\ref{sec:pathway}).

We further provide a theoretical explanation by first interpreting visual analytics as a model-building process, in which human reasoning and computational learning are tightly coupled. We then adopt an information-theoretic perspective to explain how visual analytics contributes to the optimization of machine learning workflows (Section~\ref{sec:theoretical_explain}).

In summary, the main contributions are as follows:
\begin{itemize}
    \item A structured categorization scheme for VIS4ML systems from four complementary perspectives.
    \item A set of knowledge injection pathways that characterize how human knowledge is transformed and propagated through visualization, interaction, and actions into ML workflows.
    \item A theoretical explanation of VIS4ML, interpreting the survey findings through the conceptual lens of VA as model building, and providing an information-theoretic perspective on why and how it optimizes ML workflows.
\end{itemize}

\section{Related Work}
\subsection{Machine Learning Workflow}

\textcolor{black}{Recent work from the machine-learning side increasingly shows that model development is not a one-shot act of fitting a learner, but an iterative process organized around repeated revision of data, features, models, and evaluation \cite{sculley2015hidden,amershi2019software,xin2018accelerating}. Sculley et al.\  draw attention to the broader system-level burden of ML, arguing that long-term difficulty often lies in dependencies, feedback loops, and maintenance costs surrounding the model rather than in the learning component alone\cite{sculley2015hidden}. Amershi et al. make a similar point from industrial practice, describing ML development as a workflow that spans data collection, cleaning, labeling, feature engineering, training, evaluation, deployment, and monitoring, with many feedback loops rather than a clean linear pipeline\cite{amershi2019software}. Xin et al. argue that ML is iterative by nature. It is less about a single build and more about a constant cycle of testing and refining preprocessing, features, models, and evaluation choices in response to earlier results \cite{xin2018accelerating}.  Within human-in-the-loop (HITL) research, Chai and Li broaden the role of the human beyond annotation to data extraction, integration, cleaning, iterative labeling, and model training and inference \cite{chai2020human}, while Wu et al. survey the area through data processing, interventional model training, and system construction/application \cite{WU2022364}. At the same time, this literature makes clear that effective human participation does not follow simply from placing a person somewhere in the pipeline. It depends on how expertise is extracted, interpreted, and incorporated under practical constraints such as cost, timing, interface design, trust, and system integration \cite{holzinger2016interactive,dudley2018review,gomez2024human,10530996}. From health informatics, Holzinger argues that when data are sparse, uncertain, or otherwise difficult, domain expertise can be indispensable to the discovery process \cite{holzinger2016interactive}. Current frameworks fail to systematically track the lifecycle of human knowledge within the ML workflow from its initial formalization into data to its subsequent propagation through the pipeline. VIS4ML is designed to bridge this specific gap, providing the necessary infrastructure to manage these HITL transitions. }
\subsection{Visual Analytics for Machine Learning}

\textcolor{black}{
Prior work has examined visualization and machine learning from several complementary perspectives. In VIS4ML more broadly, Tam et al.~\cite{Tam:2016:TVCG} analyzed the role of humans in ML workflows, while Sacha et al.~\cite{Sacha2019TVCG} proposed an ontology describing where visualization can support model development. Subsequently, surveys investigated the field by data perspective~\cite{Wang2024TVCG}, analysis stage~\cite{Yuan2021}, deep learning research~\cite{Hohman2019}, trust ~\cite{Chatzimparmpas2020}, human-centered ML evaluations~\cite{Sperrle2021},  and by integrating ML into visual analytics~\cite{Endert2017}. Despite the field's wide-ranging focus, multiple research gaps exist. To the best of our knowledge, no survey maps how visual analytics is used to inject human knowledge into machine learning workflows. Adjacent concepts, such as refinement~\cite{Wang2024TVCG}, steering~\cite{Yuan2021} and semantic interaction~\cite{Endert2017} are rarely the predominant organizing focus. Furthermore, a multitude of surveys cover interpretability, trust, and explainability ~\cite{Chatzimparmpas2020, Hohman2019, Sperrle2021}, rather than focusing on the actions in machine learning workflows resulting from visualizations and interactions.}

\textcolor{black}{Beyond these surveys, prior VIS4ML systems have been designed for different model families and application tasks. For example, Wang et al.~\cite{Wang:2020:TVCG} put forth the visual analytics tool HypoML and support hypothesis-focused assessments of machine learning models. Shen et al.~\cite{Shen:2020:PVIS} and Wang et al.~\cite{Wang:2021:TVCG} studied RNN-based workflows in temporal prediction settings, Liu et al.~\cite{Liu:2017:TVCG} focused on decision-tree models, and Chatzimparmpas et al.~\cite{Chatzimparmpas:2023:InfoVIS} investigated random forests. XAI-related tools facilitate model-agnostic analysis and fairness evaluation~\cite{Wexler2019}. More recently, visualization has also been used to examine modern language models and LLM-related analysis settings~\cite{Sevastjanova:2023:VDS, Strobelt2022}. These studies show that VIS4ML has been explored across different models and tasks, but prior work still lacks an account of how visualization supports human knowledge injection to improve ML systems.}

\subsection{Related VIS Concepts and Theories}
Following the above discussions, HITL is an accepted approach in ML, and interactive visualization facilitates HITL in ML. There have been different conceptual discourses for articulating the benefits of VIS \cite{Streeb:2021:TVCG,Streeb:2021:CGA,Chen:2020:book}, which can also be used to explain many aspects of VIS4ML. For example,
\emph{Insightism} (e.g., \cite{vanWijk:2005:Vis,Chang:2009:CGA}) -- ML developers may gain insight by visualizing various data in an ML workflow;
\emph{Cognitivism} (e.g., \cite{Stasko:2014:BELIV}) -- visualization may give ML developers or users more confidence about an ML model;
\emph{Communicationism} (e.g., \cite{Evergreen:2016:book}) -- visualization may facilitate explainable AI;
\emph{Economism} (\cite{Tufte:2001:book,Ware:2004:book}) -- ML developers may become more efficient when assisted by VIS;
\emph{Hypothesism} (e.g., \cite{Suh:2022:arXiv}) -- VIS can help ML developer hypothesize the optimal parameter setting; and
\emph{Pragmatism} (\cite{Marty:2009:book,Chen:2014:book}) -- ML developers and users find VIS as useful aids for performing different tasks.

Almost all these conceptual discourses focus on the benefits gained by VIS users, including ``gaining new knowledge from data'', but they do not imply directly that ``VIS users inject knowledge into ML workflows''. In this work, we interpret our survey findings through the existing conceptual model that considers VA as model building \cite{Andrienko:2018:CGF}, using it to articulate the VA's role in knowledge injection.  

The information-theoretic cost-benefit analysis proposed by Chen and Golan \cite{Chen:2016:TVCG} can be considered as a mathematical abstraction of the conceptual discourses of economism. They categorized VIS for developing ML models as well as numerical simulation models as \emph{level 4 visualization} -- the most complex class of visualization tasks.
Chen further explained the complexity of searching for an optimal model, and used cost-benefit analysis to show that human knowledge is used to reduce the search space in the ML workflows \cite{Chen:2020:bookOUP}.
Tam et al. first used information theory to estimate the amount of human knowledge that was used in two ML case studies \cite{Tam:2016:TVCG}.
In this work, we make a broad observation of a wide range of ML workflows through the lens of information-theoretic cost-benefit analysis.

\section{Survey Methodology}
\label{sec:methodology}

\begin{figure}[ht]
    \centering
    \includegraphics[width=\columnwidth]{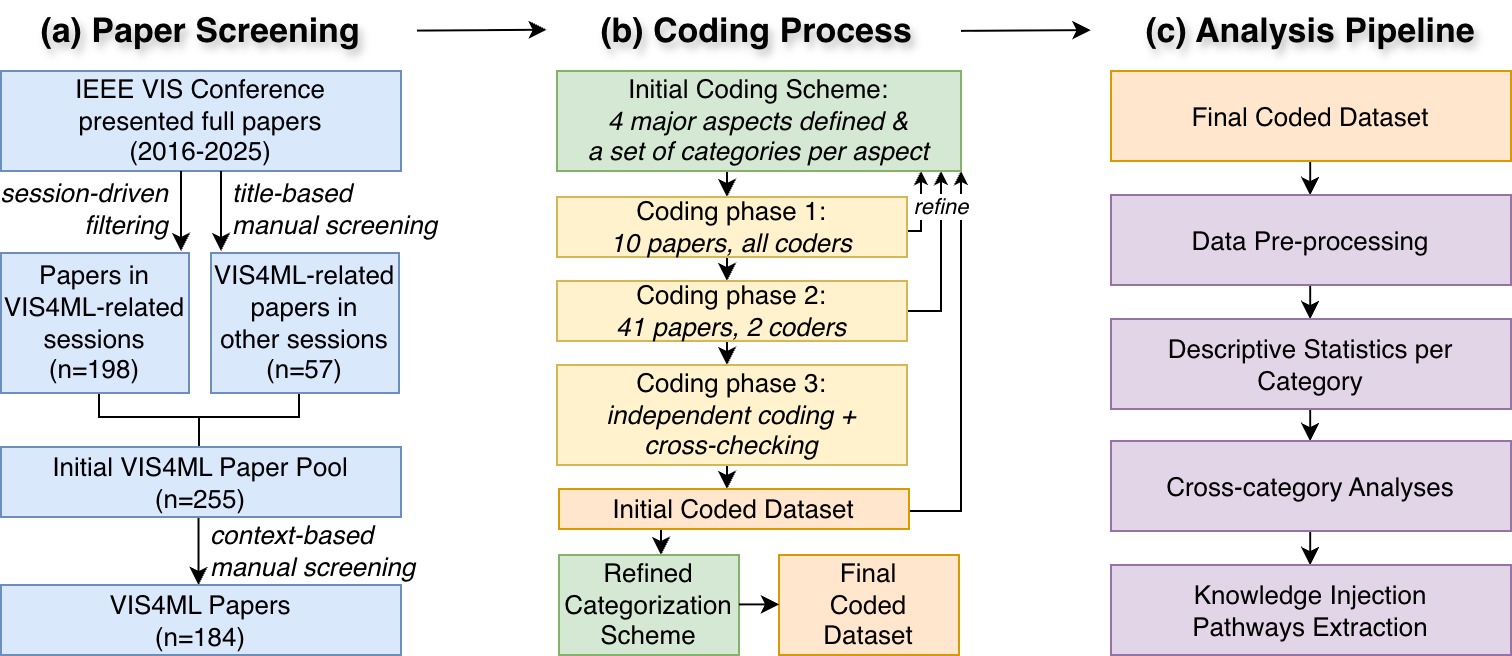}
    \caption{Overview of the survey process: (a) Paper screening; (b) Coding process and categorization scheme refinement; (c) Analysis pipeline.}
    \label{fig:SurveyMethod}
\end{figure}

This section describes the methodology used to collect, code, and analyze VIS4ML papers. The entire process is illustrated in Fig.~\ref{fig:SurveyMethod}.

\vspace{-0.3 cm}

\subsection{Paper Collection and Screening}

We collect papers from the IEEE VIS conference, the primary venue for visualization and visual analytics research. We focus on papers presented at IEEE VIS over the past decade (2016-2025), covering the rapid development of VIS4ML systems, as well as the broader transition from classical machine learning techniques to the era of generative AI (a temporal analysis of these trends is provided in Appendix \ref{app:temporal}).

To identify relevant papers, we adopt a multi-stage screening process. We first perform session-driven filtering based on session topics. Specifically, we include all papers from sessions related to VIS4ML or closely aligned themes, such as ``VA4AI'', ``Vulnerabilities in Machine Learning'', and ``Explanation, Exploration, and Model Configuration.'' Across the ten-year span, this step results in 33 selected sessions and a candidate pool of 198 papers.

To reduce the risk of missing relevant papers outside these sessions, we further conduct a title-based manual screening. This step adds 57 papers from 12 additional sessions, including best paper sessions and sessions such as ``Interactive Dimensionality'', ``Decision Making'', and ``Analytics and Reasoning,'' resulting in an initial pool of 255 papers.

We then conduct a context-based manual screening process, in which coders read each paper to assess its relevance. We exclude papers that are not directly related to VIS4ML, including those that apply ML and VA to domain-specific problem solving, where VA is not used to support ML itself. We also exclude survey, methodological, or empirical studies that do not involve concrete VA systems. In addition, we exclude works that focus solely on ML for visualization (ML4VIS) or do not involve human interaction in the ML workflow. The final dataset consists of 184 papers, which forms the basis for our subsequent coding and analysis.

\subsection{Coding Scheme}
\label{sec:coding_scheme}
To investigate how human knowledge is injected into machine learning workflows through visual analytics, we develop a structured coding scheme that captures different aspects of VA-supported ML processes. This task is inherently challenging, as human knowledge is abstract and not directly observable. Instead, it must be inferred from observable elements of VA systems and reported user behaviors.

In particular, we approximate knowledge injection through three types of evidence: (1) we analyze the \textbf{Visualization} design to understand what data, model components, or intermediate results are presented to users, which reflects what information is made available for human reasoning; (2) we examine \textbf{Interaction} mechanisms to characterize how users are allowed to explore, manipulate, or steer the ML process; (3) we identify user \textbf{Actions} reported in the papers, especially those that lead to changes in data, features, models, or training processes, which serve as concrete manifestations of human knowledge being injected into the workflow.
In addition, since human intervention is dependent on the underlying ML context, we also capture key \textbf{characteristics of the ML setup}, as different ML settings may lead to different patterns of human involvement and knowledge injection.

Based on these considerations, we organize our coding scheme into four perspectives as listed below, each consisting of multiple categories. 
Each category further contains a set of instances (i.e., coding options). Instances in categories marked with * are not mutually exclusive, which means a paper can be associated with multiple instances in such categories, whereas the remaining categories are mutually exclusive, with each paper assigned to a single instance. The complete list and detailed definitions of all instances are provided in Appendix~\ref{apx:Tables}.

\paragraph{ML Characteristics Perspective (9 Categories)} aims to capture the key factors that may influence how humans interact with machine learning models. From the ML model, we recorded its characteristics, including the employed \textit{\textbf{techniques} (1.1 ML\_TECH*)}, \textit{\textbf{training methods} (1.2 ML\_TM*)}, and \textit{\textbf{model architectures} (1.3 ML\_ARCH*}. 
\textcolor{black}{In addition, we incorporate a \textit{\textbf{grouping of model families} (1.4 ML\_GRP)} to provide a higher-level view of model characteristics, following broad distinctions commonly recognized in the machine learning literature~\cite{fernandez2014we,kotsiantis2007supervised} and further informed by standard treatments of probabilistic and deep learning models~\cite{bishop2006pattern,goodfellow2016deep}.}
We also included contextual factors that may affect the design of visualization and interaction in a VA system, such as the \textit{\textbf{ML tasks} (1.5 ML\_TASK*)} (e.g., classification, clustering, or detection), \textit{\textbf{ML workflows} (1.6 ML\_WF)} (e.g., development or deployment), the \textit{\textbf{users} (1.7 ML\_USER*)}, and the \textit{\textbf{model inputs and outputs}} data types \textit{(1.8 ML\_IN* \& 1.9 ML\_OUT*)}. 
%
\paragraph{Visualization Perspective (6 Categories)} aims to capture how visualizations are designed in VA systems to support human understanding and reasoning about data derived from the ML processes. We defined several categories as follows: 
The \textit{\textbf{ML stages}} supported by the visualization \textit{(2.1 VIS\_MLS*)}, which includes both the four stages in an ML development workflow proposed by Sacha et al.~\cite{Sacha2019TVCG}, and two more deployment workflow stages (using deployed models and third-party evaluation) proposed by us. The \textit{\textbf{visualized objects} (2.2 VIS\_OBJ*)}, for example, input data, features, model states, and performance metrics.
Rather than recording specific chart names being used in the papers, which can vary across systems, lack standardization, and often appear in combination, we characterized visualization design through the following aspects. We identified the \textit{\textbf{visualization types} (2.3 VIS\_TYP*)}, distinguishing between common statistical charts and more advanced visual representations. We also captured the \textit{\textbf{visual channels} (2.4 VIS\_CH*)} used in the system, such as geometric, optical, and relational encodings~\cite{Chen2014Springer}. The \textit{\textbf{number of visualization views} (2.6 VIS\_NoC)} was collected to reflect the structural complexity of the VA system.
We also coded the \textit{\textbf{analytical patterns} (2.5 VIS\_PTN*)} enabled by the visualization as a semantic abstraction of the insights conveyed to users, capturing what users are expected to perceive and reason about through the visualizations.

%
\paragraph{Interaction Perspective (6 Categories)} focuses on how interaction mechanisms enable users to inject knowledge back into the ML workflow, rather than only acquiring insights from it (the visualization perspective). 
Similar to the visualization part, we captured the \textit{\textbf{ML stages} (3.1 INT\_MLS*)} at which interaction is supported, indicating where users are able to engage with and influence the model. 
We then characterized \textit{\textbf{interaction types}} through four aspects: (1) navigation and visual exploration \textit{(3.2 INT\_TYP\_NAV*)}, (2) ML development workflow steering \textit{(3.3 INT\_TYP\_MLDv*)}, (3) ML deployment workflow steering \textit{(3.4 INT\_TYP\_MLDp*)}, and (4) human annotation and feedback \textit{(3.5 INT\_TYP\_AnF*)}.
Finally, we included \textit{\textbf{interaction modalities} (3.6 INT\_MOD*)} as an additional aspect to capture how interaction is implemented from a user interface (UI) perspective, such as through direct manipulation, programming interfaces, or natural language input.

%
\paragraph{Action Perspective (3 Categories)} focuses on the outcomes of user interaction, capturing how human knowledge is operationalized into actions that affect the ML workflow. These actions may be performed within the VA system or externally.
We captured the \textit{\textbf{action types} (4.1 ACT\_TYP*)} through which human knowledge is injected into the ML process, such as data editing, feature engineering, model selection, or hyperparameter tuning. These actions are represented by the ML instances that users modify or refine.
Recognizing that actions are highly dependent on how they are presented and reported in the literature, we identify \textit{\textbf{action evidence} (4.2 ACT\_EVD)} to assess the extent to which such actions are supported and realized. Specifically, we distinguished between concrete actions explicitly reported, envisioned actions suggested by the system, and cases where no action is supported or reported.
Finally, we considered \textit{\textbf{knowledge injection mechanisms beyond the VA system} (4.3 ACT\_KI*)} by capturing how users may influence the model through external means when such actions are not directly supported by the VA system, to identify gaps in current VA systems and understand where human intervention occurs outside of visual analytics support.

%
\subsection{Coding Process and Validation}
The coding process consists of two interrelated components:
1) the development of the coding scheme, including the definition of categories and their corresponding classes (Section~\ref{sec:coding_scheme}), and
2) the application of this scheme by coders to annotate the collected papers.
The process is inspired by grounded theory~\cite{Alexandra2025TVCG}, following an iterative approach. The initial scheme is progressively refined through empirical coding.

All authors of this paper were involved in the initial brainstorming to establish the coding scheme. Through collaborative discussions, we developed a preliminary categorization framework with four major aspects. Building on the framework, the first six authors conducted the coding and iterative refinement process. We began with a calibration phase by selecting 10 papers. Each coder independently applied the current coding scheme to these papers, followed by group discussions in which discrepancies were examined and resolved. This process was repeated over three rounds of discussion and calibration, during which both the high-level categories and fine-grained classes were refined, and a shared understanding among coders was established.

Then, 15\% papers (n=41) were selected for two-coder independent coding. The results were then compared to assess inter-coder agreement (agreement rate=94.21\%, across 41 papers and 24 categories; statistics per category are provided in supplementary materials), followed by discussions to reconcile differences. 

The remaining papers were randomly assigned to the six coders for individual coding. To ensure the reliability, a second coder reviewed cases where the primary coder had uncertainties, and an additional subset of papers was randomly sampled for cross-checking.

After completing the first round of coding, all authors jointly reviewed and discussed the aggregated results. Guided by insights from the data, we further refined the coding scheme by merging categories with overlapping meanings (e.g., \textit{Data Sample Adjust} and \textit{Data Structure Adjust} into \textit{Data Adjust} in \textit{Interaction} part) and splitting overly broad categories with high coverage into more fine-grained classes (e.g., \textit{Model Adjust} in \textit{Interaction} and \textit{Structured Input} in \textit{Action}). We then recoded the affected papers to align with the updated scheme. This iterative process led to a finalized coding scheme and a set of coding results agreed upon by all authors.

\subsection{Analysis Pipeline}
To extract knowledge injection patterns from the coded dataset, we first conducted descriptive analysis by summarizing the proportion of papers associated with each instance within a category. 
We then explored relationships across categories by analyzing co-occurrence patterns, such as the connections between \textit{ML Techniques} and \textit{VIS Types}, and between \textit{ML Tasks} and \textit{Interaction or Action Types}. We further performed topic modelling to explore the association patterns across multiple categories (Section~\ref{sec:coding_results}). Based on these cross-category relationships and contextual reference to the original papers, we extracted representative knowledge injection pathways, revealing how different components are combined in practice (Section~\ref{sec:pathway}).

\section{Coding Results}
\label{sec:coding_results}

\subsection{General Statistics}
\label{sec:generalstat}
Based on the coding results, our analysis reveals several general patterns shared across VIS4ML systems. 
Since a single paper may be associated with multiple instances within each category, 
the reported percentages for these categories are calculated as the number of papers assigned to each instance divided by the total number of papers. These instance-level patterns for all the 24 categories can be observed in Figs.~\ref{fig:Sankeys1} and \ref{fig:Sankeys2} by examining the source and target columns of the Sankey diagrams.

From the ML characteristics, 
neural network–based models dominate the literature (Fig.~\ref{fig:Sankeys1} (a) source column). 24.5\% of the papers are associated with convolutional neural networks (CNNs), followed by large language models and transformers (19.6\%). 
In terms of tasks (Fig.~\ref{fig:Sankeys1} (e) source column), classification is by far the most common (43.5\%), with natural language processing (21.7\%) and clustering (17.4\%) also widely represented. Most systems focus on ML development workflows (64.1\%), whereas fewer works address deployment workflows (20.1\%) or both (15.8\%).

In terms of visualization, as illustrated in the target column in Fig.~\ref{fig:Sankeys1} (c),
common statistical charts and multivariate tabular visualizations are prevalent, appearing in 69\% and 60.9\% of the papers, respectively. More complex representations, such as graph-based, hierarchical, and temporal visualizations, are used less frequently. 
At the same time, navigation-based interactions are implemented in all surveyed VA systems for local data exploration. Specifically, as shown in the target column in Fig.~\ref{fig:Sankeys2} (g), selection/focus (90.2\%), details-on-demand (66.3\%), and filtering (58.7\%) are the most frequently supported operations, and they are not mutually exclusive. 
Taken together, visualization and interaction form the core components of VA. The consistent use of standard visualizations with local data exploration interactions suggests that they serve as a foundational design pattern across the surveyed VIS4ML systems.

To understand knowledge injection pathways, we analyze how user actions are reported across the surveyed papers, as actions serve as observable representations of human knowledge (captured by 4.1 ACT\_EVD). 
As illustrated in the target column in Fig.~\ref{fig:Sankeys2} (k), 60.3\% report concrete actions supported by VA systems, indicating that VIS4ML research has made substantial progress in enabling users to actively intervene in and improve ML processes. However, a notable portion of works (15.2\%) do not report any action, focusing primarily on model understanding, as commonly seen in explainable AI systems. This raises an important question: while such systems help users understand the model, how this understanding is subsequently translated into actionable improvements remains unclear.
In addition, some works describe only envisioned actions (12\%), or report actions performed outside the VA system (12.5\%), such as offline model development or programming-based workflows. This suggests that certain forms of knowledge injection are not yet fully supported within current VA systems. On one hand, this reveals opportunities for extending VA to better support these external actions. On the other hand, it may also reflect practitioners’ preferences for alternative interaction modalities, such as programming interfaces, when performing certain types of model modifications.


%
\subsection{Pairwise Co-occurrence Analysis}

To investigate relationships between categories, we conducted a pairwise co-occurrence analysis to capture how classes in two categories are related, which forms a fundamental basis for constructing part of the knowledge injection pathways.
The results are visualized using Sankey diagrams, which encode both the number of papers associated with each instance and the co-occurrence flows between categories. Specifically, the width of each node reflects how many papers contribute to the corresponding instance on the source or target side, while the links between nodes indicate the frequency of co-occurrence between instances from the two categories. It is important to note that, for categories with instances that are not mutually exclusive, the total width of nodes may vary because a single paper can contribute to multiple instances.
When considered independently, each side of the Sankey diagram (source or target) reflects the number of papers associated with each instance, providing a visual counterpart to the instance-level statistics reported in the previous section (Section~\ref{sec:generalstat}).

\begin{figure*}[ht]
  \centering
  \begin{tabular}{@{}c@{}c@{}c@{}}
    
    \includegraphics[width=60mm]{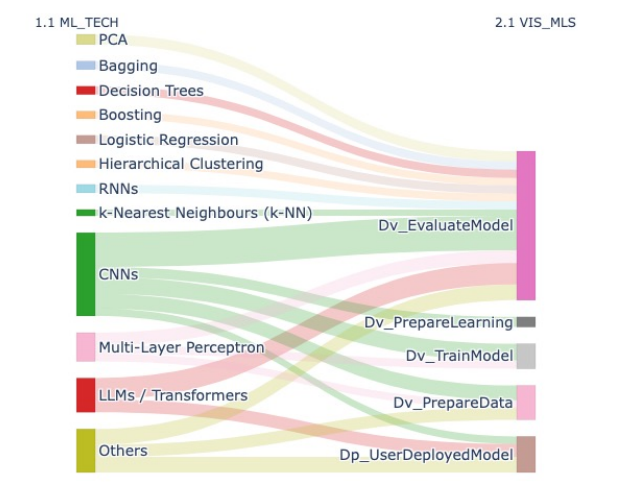} &
    \includegraphics[width=60mm]{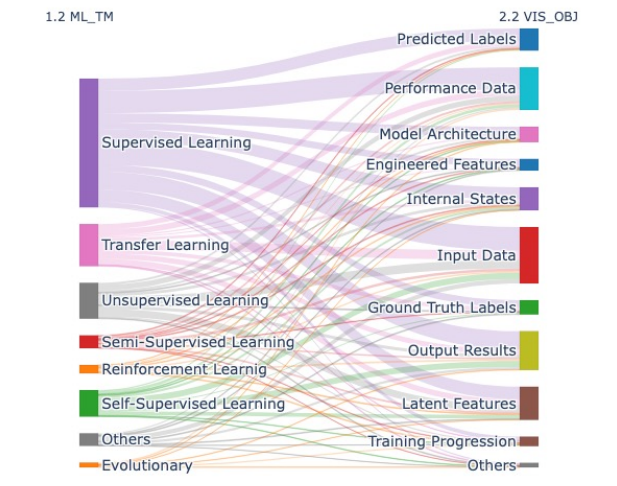} &
    \includegraphics[width=60mm]{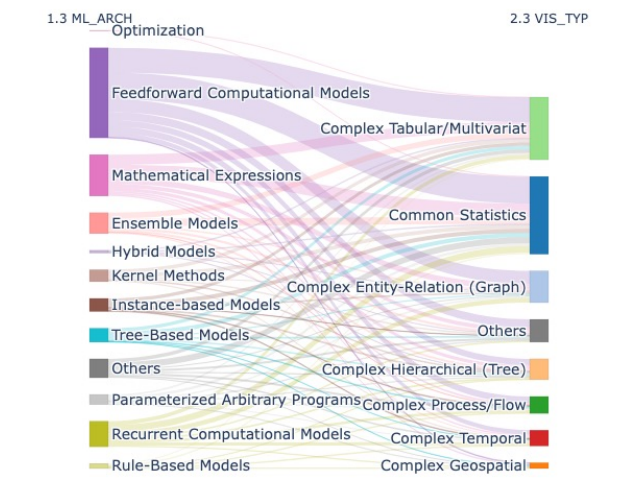} \\
    (a) ML\_TECH (top 12) vs. VIS\_MLS &
    (b) ML\_TM vs. VIS\_OBJ &
    (c) ML\_ARCH vs. VIS\_TYP \\[1mm]

    \includegraphics[width=60mm]{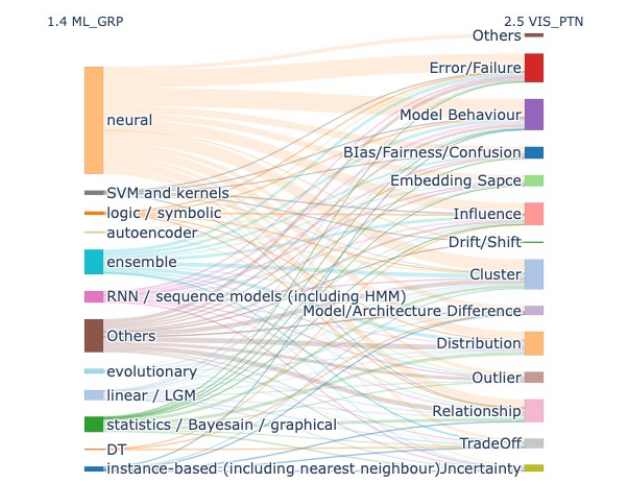} &
    \includegraphics[width=60mm]{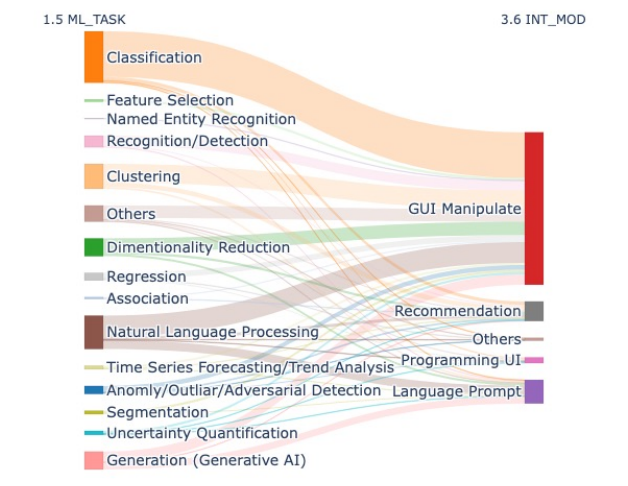} &
    \includegraphics[width=60mm]{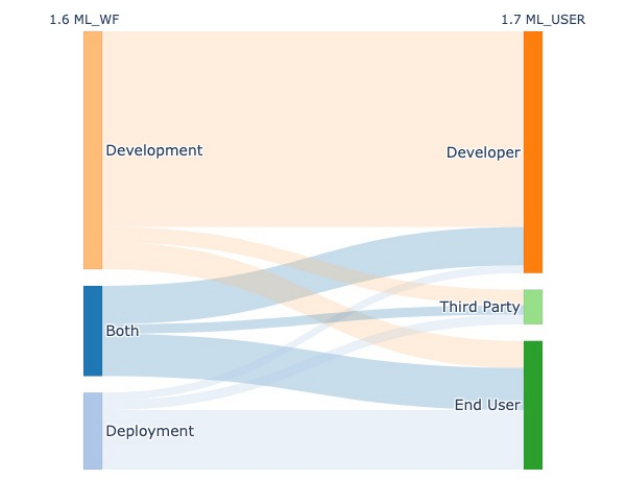} \\
    (d) ML\_GRP vs. VIS\_PTN &
    (e) ML\_TASK vs. INT\_MOD &
    (f) ML\_WF vs. ML\_USER \\[-2mm]


  \end{tabular}
  \caption{Selected pairwise Sankey diagrams (part 1 of 2).}
  \label{fig:Sankeys1}
  \vspace{-0.3 cm}
\end{figure*}

\begin{figure*}[ht]
  \centering
  \begin{tabular}{@{}c@{}c@{}c@{}}
    


    \includegraphics[width=60mm]{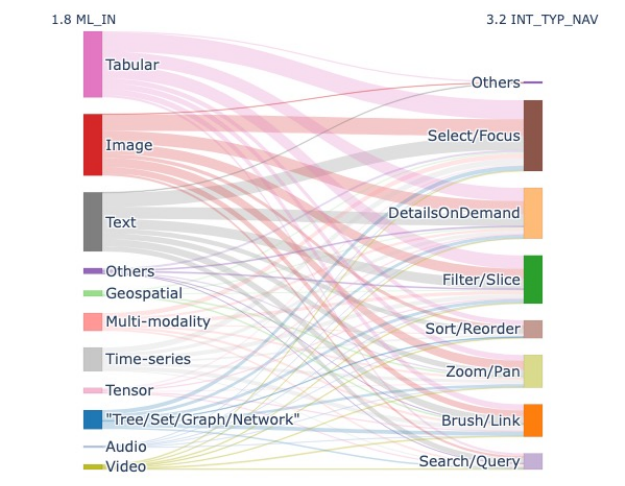} &
    \includegraphics[width=60mm]{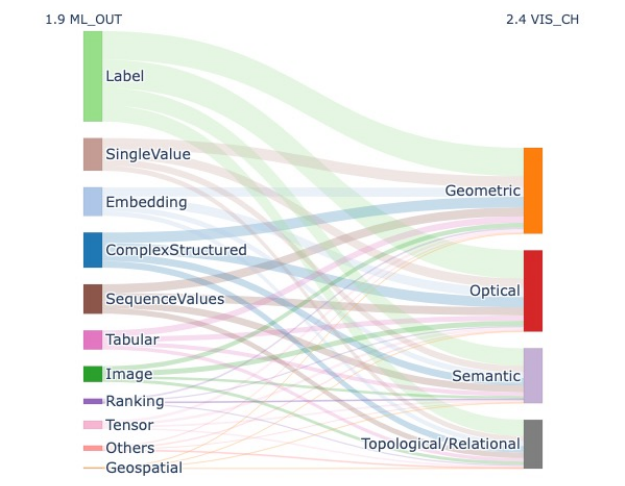} &
    \includegraphics[width=60mm]{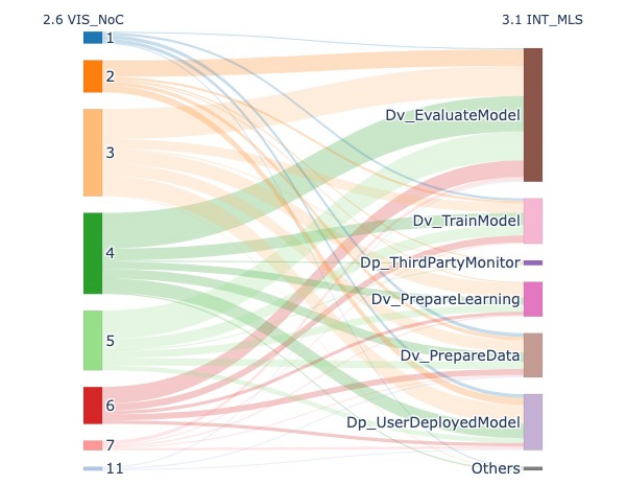} \\
    (g) ML\_IN vs. INT\_TYP\_NAV &
    (h) ML\_OUT vs. VIS\_CH &
    (i) VIS\_NoC vs. INT\_MLS \\[1mm]

    \includegraphics[width=60mm]{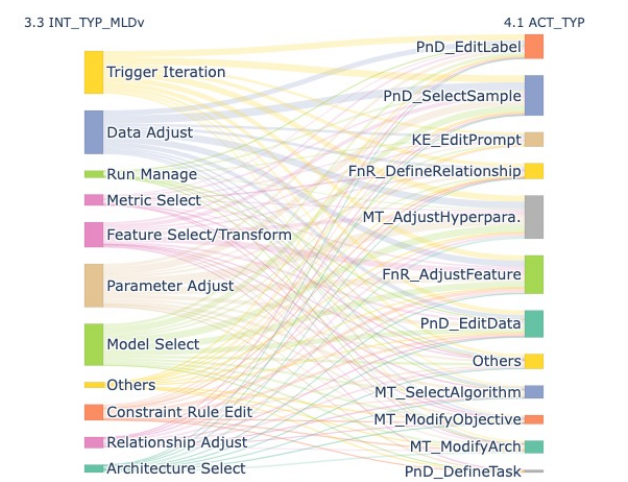} &
    \includegraphics[width=60mm]{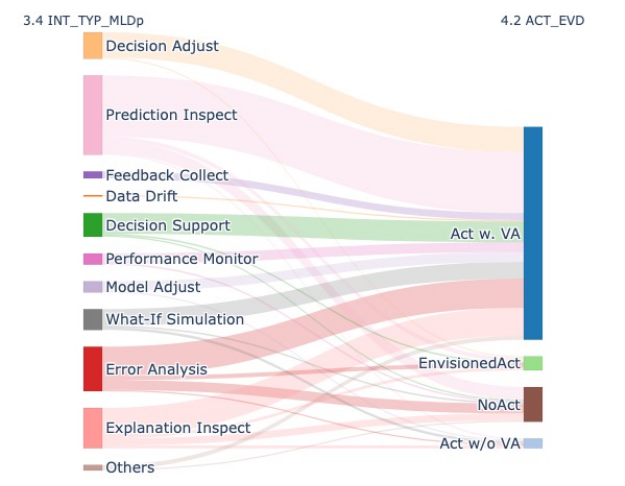} &
    \includegraphics[width=60mm]{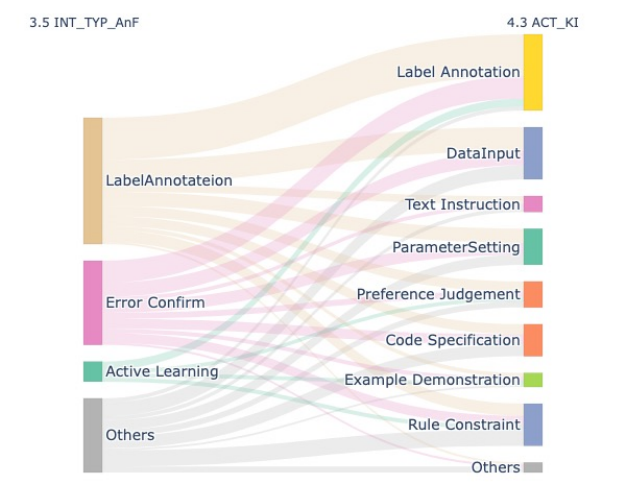} \\
    (j) INT\_TYP\_MLDv vs. ACT\_TYP &
    (k) INT\_TYP\_MLDp vs. ACT\_EVD &
    (l) INT\_TYP\_AnF vs. ACT\_KI \\[-2mm]

  \end{tabular}
  \caption{Selected pairwise Sankey diagrams (part 2 of 2).}
  \label{fig:Sankeys2}
  \vspace{-0.3 cm}
\end{figure*}

We carefully selected 12 representative pairwise Sankey diagrams for inclusion in this paper, illustrated in Fig.~\ref{fig:Sankeys1} and Fig.~\ref{fig:Sankeys2}. Together, these diagrams cover all 24 categories in our coding scheme across the four perspectives (ML, VIS, INT, and ACT), while highlighting association patterns that are particularly informative for interpreting knowledge injection pathways. The pairwise analysis could, in principle, examine 24 $\times$ 23 $\div$ 2 directed category combinations, and the full set of results is provided on the supplementary material website. 

In practice, some category pairs are more revealing than others. For example, the association between ML categories and VIS categories is especially informative because visualization is often used to help humans understand ML models, whereas the relationship between INT categories and ACT categories directly reflects how users' interaction is translated to concrete modifications of the ML workflow. Such pairs, therefore, provide focal points for our analysis.
The following presents selected findings from the analysis of these category pairs.

Fig.~\ref{fig:Sankeys1} (e) shows the co-occurrence between ML tasks and interaction modalities. As expected, GUI-based manipulation dominates across all tasks, reflected by the consistently strong flows connecting each task to this modality. Beyond this general pattern, the figure reveals a strong association between NLP-related tasks and language prompts, with a similar pattern observed for generation tasks. This suggests a knowledge injection pathway in which users interact with deployed models through natural language and iteratively adjust inputs to obtain improved outputs. In this context, language prompts serve as a mechanism for such adjustments, enabling users to refine model behavior without modifying the underlying model. This pathway is particularly prominent in VA systems for tasks related to NLP and generative AI.


Fig.~\ref{fig:Sankeys2} (j) visualizes the relationship between interaction types for ML development steering and the corresponding action types. The results indicate that many systems support interaction functionalities, such as triggering iterations, selecting models, adjusting data, and tuning parameters.
Mapping these interaction functionalities to the reported user actions reveals strong alignments between interaction types and their corresponding actions. For instance, feature selection interactions are associated with feature adjustment actions, while data adjustment interactions correspond to actions such as data or label editing and sample selection. 
Interestingly, parameter adjustment interactions show the strongest association with hyperparameter tuning actions, as reflected by the thickest flow in the diagram. This suggests a prominent knowledge injection pathway in which users iteratively adjust model parameters and hyperparameters to optimize model performance.
 

Beyond the two patterns discussed above, we also observe unexpected associations when analyzing the relationship between ML architectures and VIS types (Fig.~\ref{fig:Sankeys1} (c)): 
1) Tree-based models are not as frequently associated with tree-based visualizations as we might expect. This suggests that the choice of visualization is not always aligned with the inherent structure of the underlying model, and that alternative representations are often preferred in practice; 
2) Complex process or flow visualizations, which could support a holistic view of iterative ML workflows, are not widely adopted. Despite their potential to convey the overall modeling process, such visualizations remain relatively underexplored across VIS4ML systems.

Across categories related to visualization, see Fig.~\ref{fig:Sankeys1} (b,d), we observed relatively balanced distributions of instances in \textit{objects being visualized} and \textit{patterns being discovered by users}. This suggests a positive trend that visualizations used in VIS4ML systems tend to provide diverse forms of support for different types of ML processes, and visualizations are designed as flexible and adaptive components, capable of accommodating a wide range of tasks, models, and workflow stages.

From the perspective of ML stages supported by visualization and interaction, both VIS and INT components are predominantly concentrated in the model evaluation stage, as shown in Figs.~\ref{fig:Sankeys1} (a) and \ref{fig:Sankeys2} (i). This pattern aligns with expectations, as the evaluation stage typically produces abundant derived data, such as performance metrics and model outputs, which provide rich information for human analysis and interpretation.
However, this concentration also reveals a potential imbalance. Compared to the evaluation stage, earlier stages of the ML workflow, such as data preparation, learning configuration, and model training, receive relatively less support from VA systems, which highlights an opportunity for future work to better support human involvement across a broader range of ML stages. 

\subsection{Multi-dimensional Pattern Analysis via Topic Modeling}
While pairwise analysis reveals relationships between two categories, it is insufficient to capture the more complex, multi-dimensional patterns present in VIS4ML systems. In practice, these systems involve the joint configuration of multiple components across different perspectives, including ML, VIS, Interaction, and Action.
To better explore knowledge injection pathways from a holistic perspective, we therefore seek a method that can capture such cross-perspective patterns. Given the structured coding results, where each paper is represented as a combination of categorical labels, and their inherent textual-like characteristics, we adopt a topic modeling approach to identify recurring multi-dimensional patterns.

To prepare the coded data for topic modeling, we transformed the structured annotations into a bag-of-words representation. Each coded paper contained values across hierarchical coding dimensions (e.g., columns ``1.1 ML\_TECH'', ``2.3 VIS\_TYP''), where values could be single items or comma-separated lists. We encoded each value by combining the column prefix with the item text to create unique term identifiers. For instance, a value ``2. Unsupervised Learning'' in column ``1.2 ML\_TM'' became the term \texttt{1\_2\_2\_Unsupervised\_Learning}. During encoding, we normalized all terms by replacing dots, spaces, hyphens, parentheses, and other special characters with underscores to ensure consistent tokenization. We grouped the encoded terms into four separate document fields corresponding to four major perspectives: ML characteristics (1.*), VIS (2.*), interaction (3.*), and action (4.*). This encoding preserves the hierarchical structure of our coding scheme within term identifiers while enabling standard topic modeling techniques. We applied Non-negative Matrix Factorization (NMF) to each group and to the combined coding dimensions, with each paper represented as a mixture of topics derived from its coded attributes.


We chose Non-negative Matrix Factorization (NMF) over LDA~\cite{OCallaghan-2015} because our bag-of-words representations are short and sparse - a setting in which NMF's non-negative, parts-based decompositions yield more coherent and directly interpretable topics. To determine the appropriate granularity, we iteratively fitted NMF models targeting 5 to 21 topics and evaluated each using multiple standard coherence metrics alongside visual inspection of topic distributions in 1D~\cite{tm1D} and 2D~\cite{tm2D} embeddings. Based on this analysis, we selected 12 topics as an optimal balance between interpretability and granularity.

\begin{figure}[ht]
  \centering
  \begin{tabular}{@{}c@{}c@{}}

    \includegraphics[width=45mm]{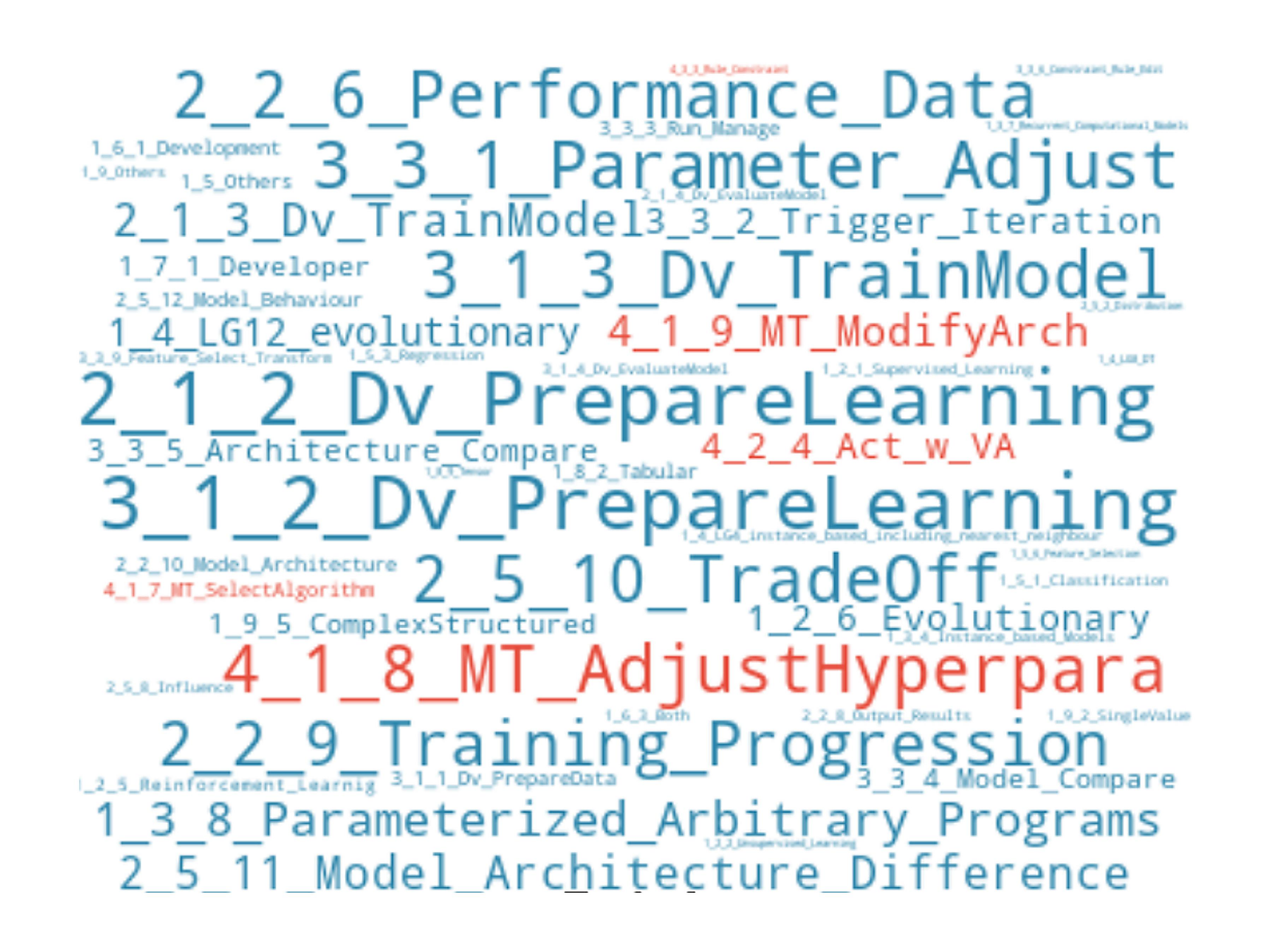} &
    \includegraphics[width=45mm]{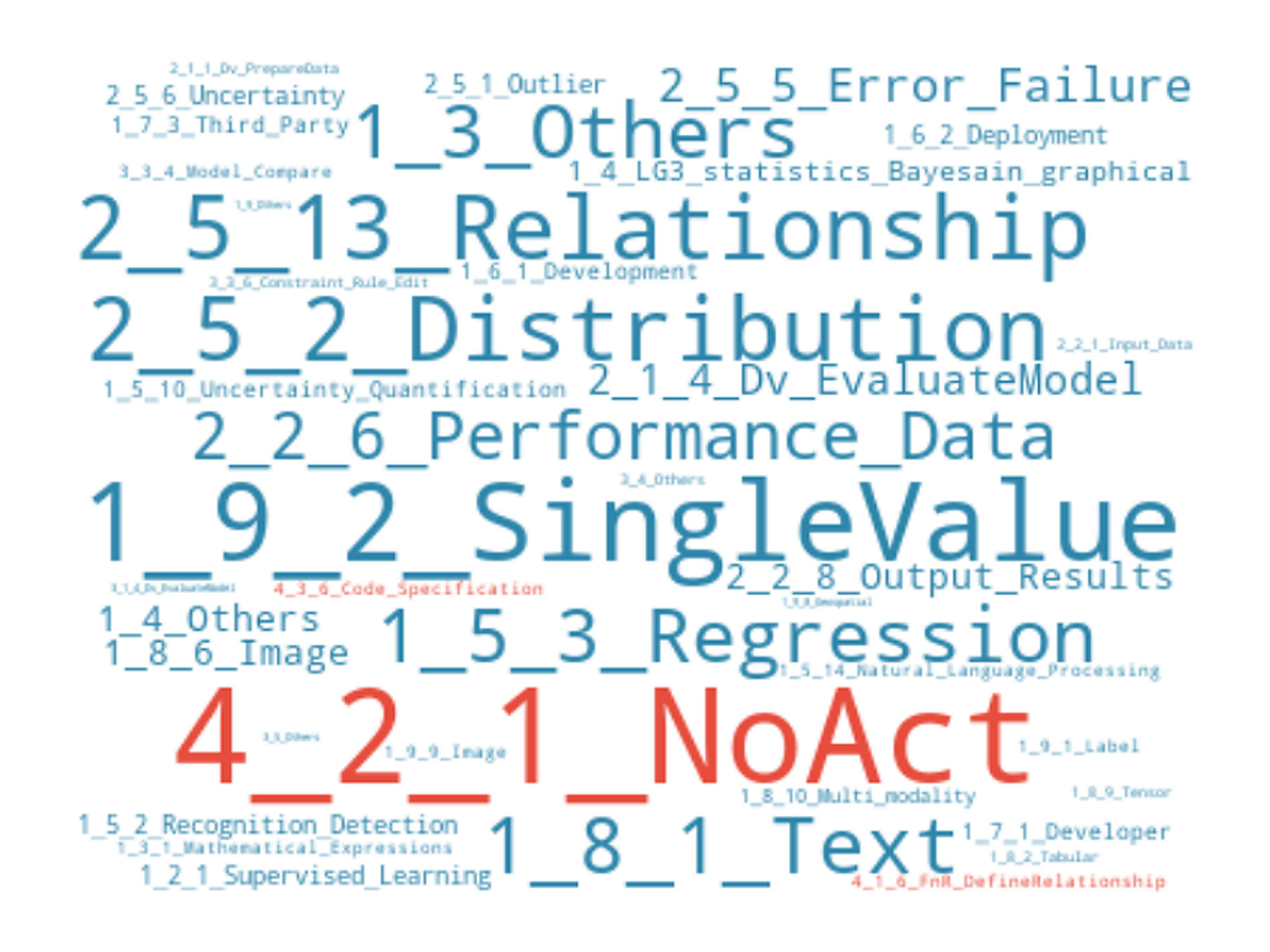} \\
    (a) Topic 0 &
    (b) Topic 4 \\

    \includegraphics[width=45mm]{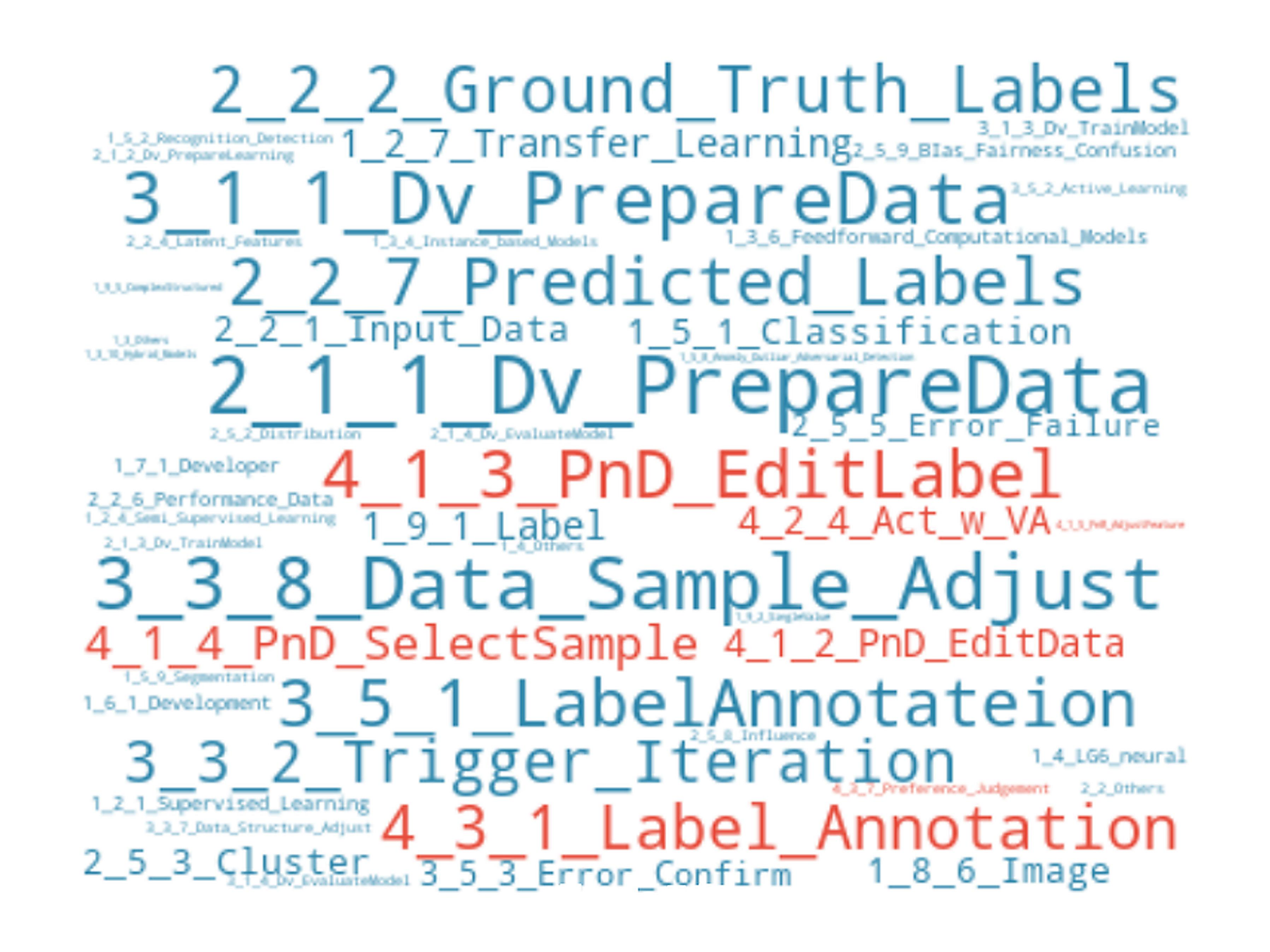} &
    \includegraphics[width=45mm]{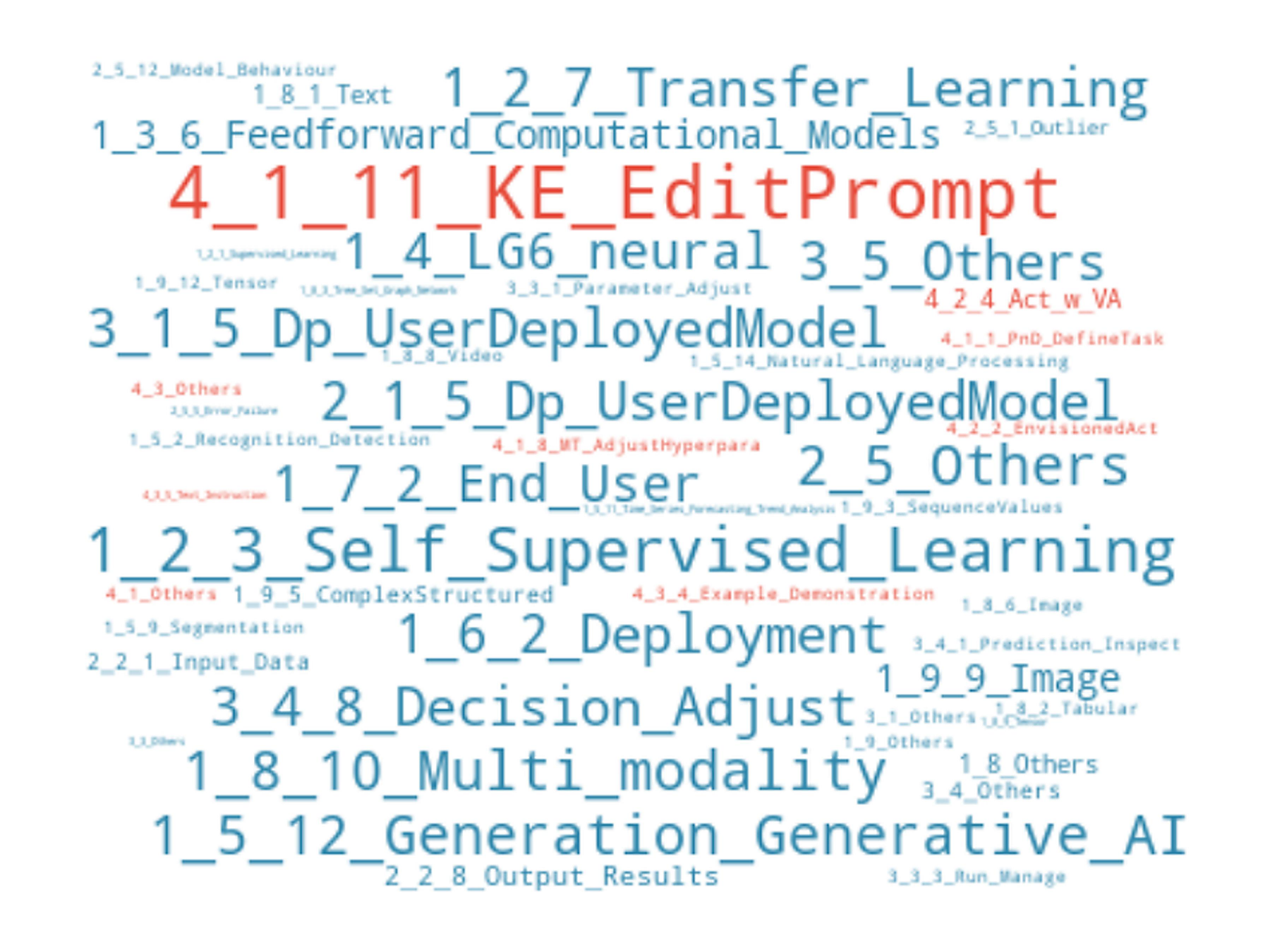} \\
    (c) Topic 6 &
    (d) Topic 11 \\[-1mm]
    
  \end{tabular}
  \caption{Word clouds of important terms in four selected topics supporting the pathway extraction in Section~\ref{sec:pathway}, chosen from a total of 12 topics. Terms from category 4 (actions) are highlighted in red, while others are in blue. Results for the remaining eight topics are provided on the supplementary website.}
  \label{fig:topics}
  \vspace{-0.3 cm}
\end{figure}

NMF produces two matrices: a document-topic matrix assigning topic weights to each paper, and a topic-term matrix capturing term importance within each topic. We visualized the topic-term matrix as a panel of word clouds (Fig.~\ref{fig:topics}), where each cloud displays the most prominent terms for one topic. Since our primary goal is understanding knowledge injection pathways, terms from category~4 (actions) are highlighted in red, while other terms appear in blue. This visualization enables rapid identification of which action types co-occur with specific ML techniques, visualization designs, and interaction patterns. 


For instance, Topic 0 in Fig.~\ref{fig:topics} (a) is centered around the action of \textit{hyperparameter adjustment} as a primary form of knowledge injection. 
Alongside this action, key terms highlight interaction mechanisms that support \textit{learning preparation} and \textit{model training}, particularly those enabling \textit{parameter adjustment}. This observation is consistent with the strong interaction--action alignment identified in our pairwise analysis.
The associated visualization components further reveal how such actions are supported in practice. In particular, systems in this topic frequently include visualizations for exploring \textit{performance metrics} and \textit{training progression}, as well as for comparing model architectures and analyzing trade-offs.
This pattern is often observed in ML workflows involving \textit{parameterized models}, such as \textit{evolutionary algorithms}, where iterative tuning plays a central role.

\section{Knowledge Injection Pathways}
\label{sec:pathway}

\begin{figure*}[ht]
  \centering
  \begin{tabular}{@{}c@{\hspace{8mm}}c@{}}
    
    \includegraphics[width=86mm]{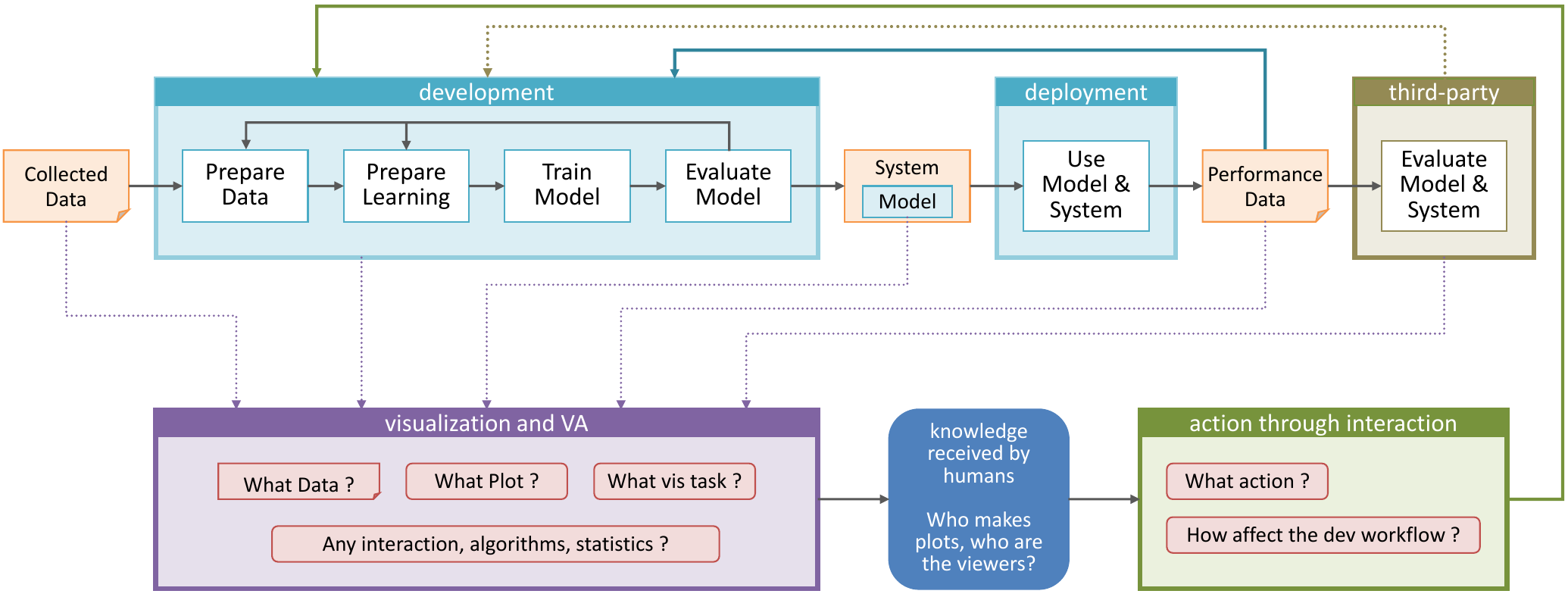} &
    \includegraphics[width=86mm]{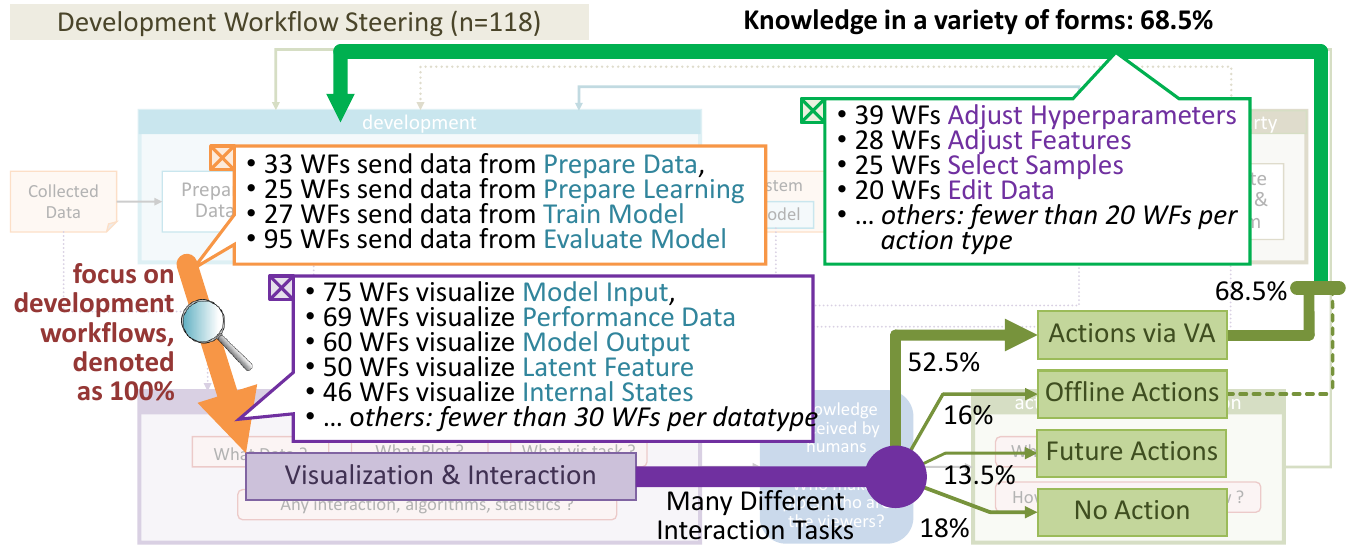} \\
    (a) ML workflow with possible knowledge injection pathways &
    (b) VA for steering development workflows \\[2mm]
    \includegraphics[width=86mm]{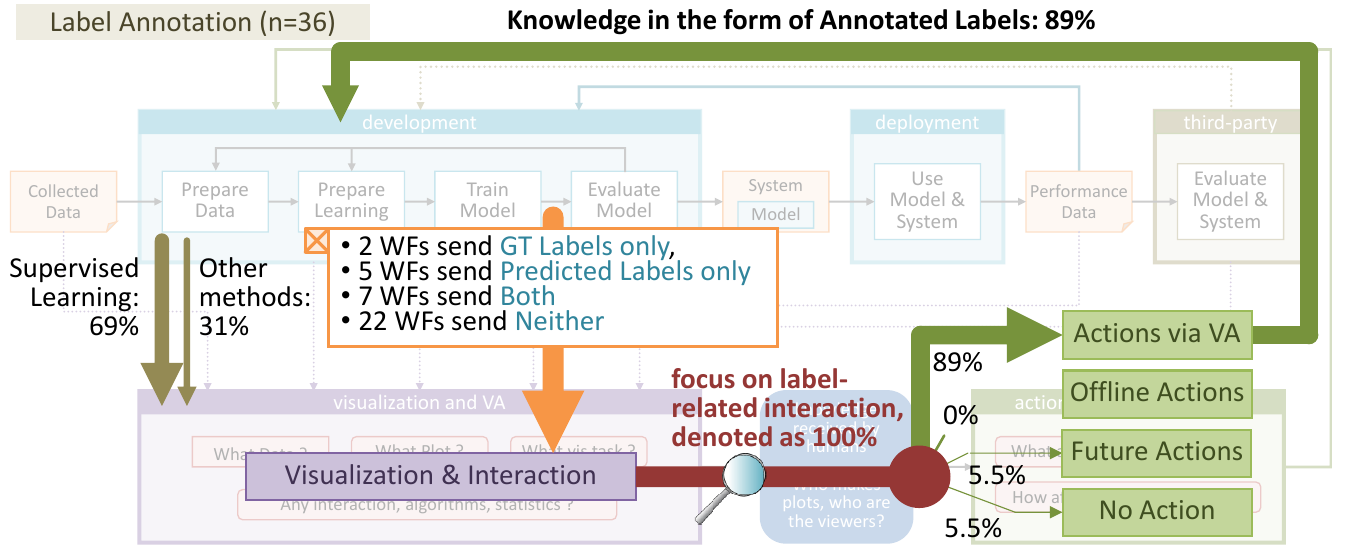} &
    \includegraphics[width=86mm]{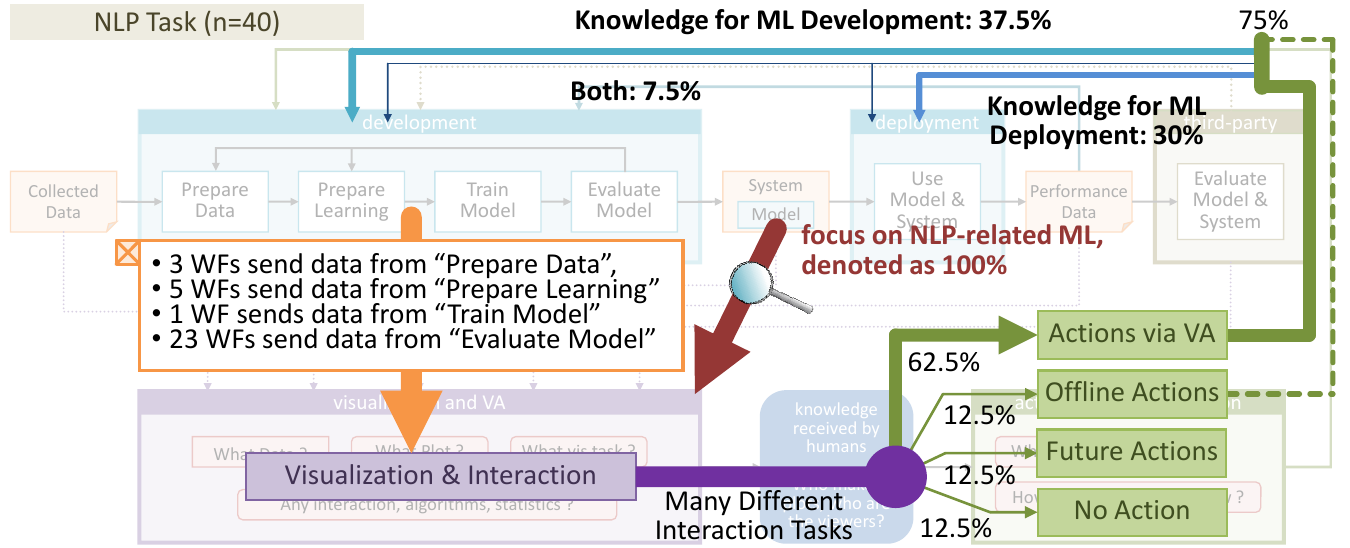} \\
    (c) VA for label annotation &
    (d) VA for supporting NLP tasks \\[2mm]
    \includegraphics[width=86mm]{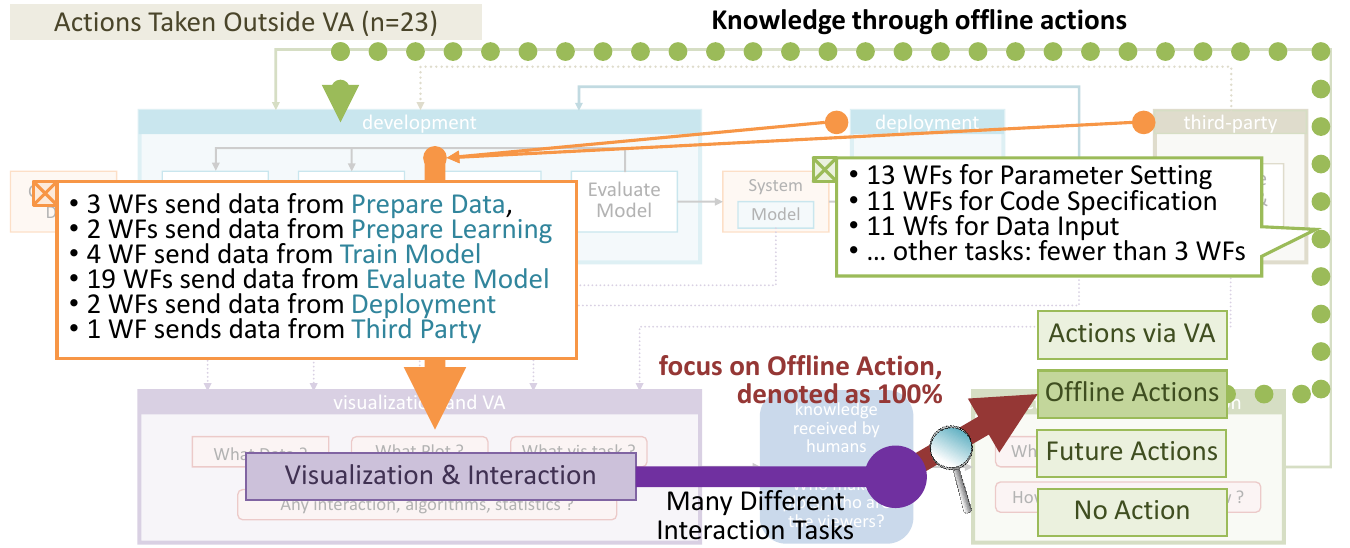} &
    \includegraphics[width=86mm]{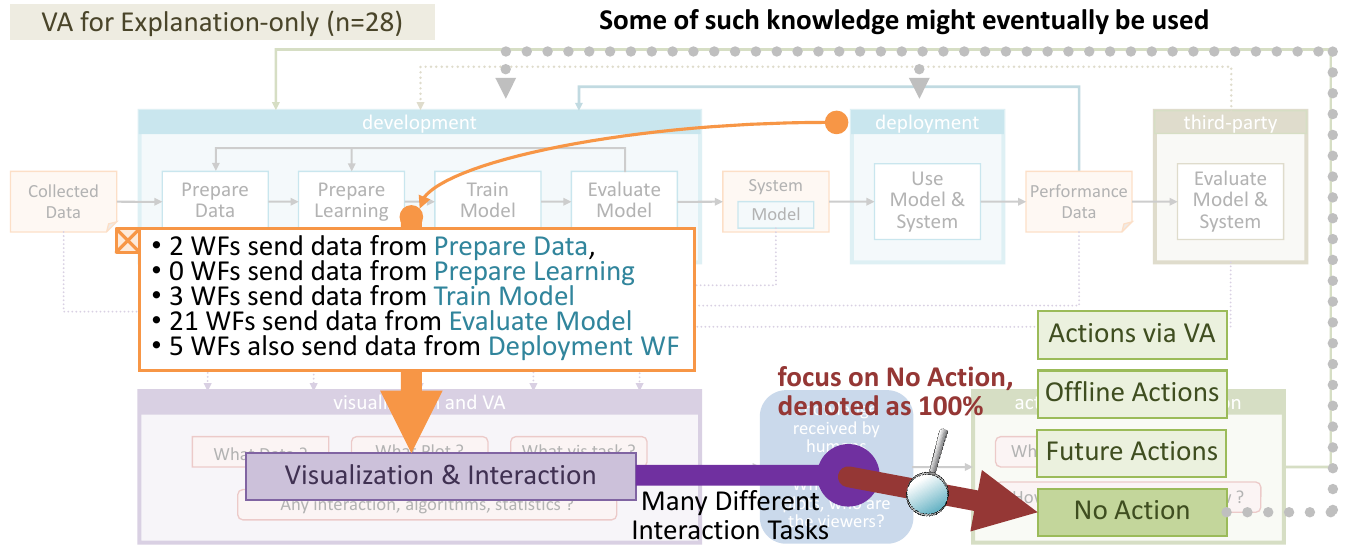} \\
    (e) Actions taken outside VA &
    (f) VA for explanation only \\
  \end{tabular}
  \caption{The survey allows us to answer questions about knowledge may be injected into ML workflows. (a) possible injection pathways, and (b)-(f) five example pathways through which human knowledge is injected into ML workflows.}
  \label{fig:Pathways}
\end{figure*}

Building upon the previous analyses, we formalize the concept of \textit{knowledge injection pathways} to describe how human knowledge is generated, transformed, and fed back into machine learning workflows through visual analytics.
Machine learning steering is inherently iterative, and thus knowledge injection does not follow a single fixed entry point. 
Fig.~\ref{fig:Pathways} (a) illustrates a general representation of knowledge injection pathways in VIS4ML systems. The figure integrates the ML workflow with visualization, interaction, and action components, providing a unified view of how human knowledge can be introduced and propagated within the process.
The upper part depicts the ML workflows, including both development and deployment stages, as well as third-party evaluation. The raw data and model outputs generated at different stages are often not directly interpretable by users. Through visualization and visual analytics, shown in the lower part of the figure and corresponding to the Visualization and Interaction perspectives in our coding scheme, these data are transformed into representations that support human understanding and insight generation. Through these components, users acquire insights and interact with the system, which enables them to perform actions that affect various parts of the ML workflow.
As such, the figure represents the space of all possible knowledge injection pathways, rather than a single predefined flow. The pathways discussed below correspond to representative patterns derived from this general structure. In the following, we illustrate several such pathways identified from our analysis, reflecting commonly observed patterns in the data.

Before presenting the pathways, we clarify the meaning of several terms used in the figures and statistics.
We distinguish between \textit{papers} and \textit{workflows}. A single VIS4ML paper may contain multiple workflow instances, each determined by a specific pathway through the four components in our coding scheme. For example, a paper may involve one workflow centered on \textit{prepare learning} and another centered on \textit{model evaluation}, where each workflow is associated with different derived data, interaction support, and resulting actions. In this sense, workflows are nested within papers and are determined by the concrete pathway structure observed in the system description and usage scenario.
In reporting statistical results, we use both percentages and counts depending on the type of categories. Percentages are computed for mutually exclusive categories, where each paper is assigned to exactly one class (e.g., action evidence). In contrast, counts are used for non-mutually exclusive categories, where a paper may belong to multiple instances (e.g., category of VIS\_TYP or INT\_TYP). Therefore, counts across such categories do not sum to the total number of papers.

%


%

\paragraph{Pathway for Development Workflow Steering.}
Figure~\ref{fig:Pathways} (b) illustrates a dominant knowledge injection pathway in which visual analytics supports the steering of ML development workflows, accounting for 80\% (n=118) of the surveyed systems. 

In this pathway, data from multiple stages of the development workflow are made available for visualization, with a particularly strong emphasis on outputs from model evaluation (95 workflows), followed by data preparation, learning configuration, and training stages. These data are transformed into visual representations that allow users to inspect various aspects of the model, including model input, performance data, model output, latent features, and internal states.

Through VA, users acquire insights and perform a variety of interaction tasks, which are then translated into concrete actions. The most common actions include adjusting hyperparameters (39 workflows), modifying features (28), selecting samples (25), and editing data (20), indicating that knowledge is injected in diverse forms.

Most of these actions are performed directly through VA systems (52.5\%), while a smaller portion occurs outside the system (16\%) or is only envisioned (13.5\%). A minority of systems (18\%) do not support explicit actions. Overall, this represents the most prevalent knowledge injection pathway in VIS4ML systems, where visual analytics is used to iteratively steer and refine models during the development workflow.

\paragraph{Pathway for Label Annotation.}
Figure~\ref{fig:Pathways} (c) illustrates a more specialized knowledge injection pathway centered on label annotation, observed in a subset of VIS4ML systems that support label-related interactions (n=36). In this pathway, human knowledge is primarily injected in the form of annotated labels.

From the ML perspective, these systems are predominantly associated with supervised learning settings (69\%), where labeled data serves as a critical component for model training.  
Beyond supervised learning, we further examine other learning paradigms that also involve label-related knowledge injection. Our coding results show that transfer learning and self-supervised learning can incorporate annotation signals in certain cases~\cite{Park2022TVCG}. In addition, label-based interaction is occasionally observed even in unsupervised settings. For example, in SOMFlow~\cite{Sacha2017TVCG}, analysts interact with self-organizing map (SOM) cells by assigning manual labels to mark subsets as interesting or uninteresting. 

As the pathway is characterized by strong VA supports in label-related information, visualizations are used to expose ground-truth labels, predicted labels, or both to users. Specifically, 2 workflows present ground-truth labels only, 5 present predicted labels only, 7 present both, and 22 do not present label information on the label side, reflecting different levels of label support in the interface.

Through visualization and interaction, users examine data instances and labels, and inject knowledge by assigning or correcting labels. These actions are overwhelmingly supported within VA systems, with 89\% of the workflows enabling actions directly through the VA interface, and only a small fraction involve envisioned or no actions (5.5\% each).
Overall, this pathway represents a supervision-driven form of knowledge injection, in which human knowledge is explicitly encoded as labeled data and directly integrated into the training process. Compared to development workflow steering, this pathway is more structured and constrained, with a strong reliance on VA-supported annotation mechanisms.

\paragraph{Pathway for Supporting NLP Tasks.}

Figure~\ref{fig:Pathways} (d) illustrates a task-driven knowledge injection pathway centered on NLP-related systems. This pathway is constructed from a subset of 40 papers in our survey that are associated with NLP tasks, representing a distinct group of VIS4ML systems within the overall corpus. 

From the ML perspective, all workflows in this pathway are associated with NLP-related tasks. Unlike development workflow steering, where knowledge injection is primarily concentrated in the development stage, this pathway spans both development (37.5\%) and deployment (30\%), with a small portion covering both stages (7.5\%). 

These observations further reveal two distinct modes within this pathway. The first mode is development-oriented, resembling the previously identified development workflow steering pathway but with a stronger emphasis on NLP tasks. In this setting, visualization primarily supports understanding and interpretation, with systems frequently presenting training data, model outputs, and intermediate representations to help users explain model behavior.
The second mode is deployment-oriented, where knowledge injection occurs during the use of deployed models. In this setting, users interact with the system by adjusting inputs, often through language-based interaction such as prompting, to iteratively refine model outputs. This mode highlights a more direct and flexible form of knowledge injection, where user feedback is incorporated at the input level without modifying the underlying model.

In general, all these interactions are translated into actions both within and beyond the VA system. Specifically, 62.5\% of workflows support actions directly through VA systems, while 12.5\% involve offline actions, and another 12.5\% describe envisioned or no actions.

\paragraph{Pathway for Actions Taken outside VA.}

Figure~\ref{fig:Pathways} (e) illustrates the knowledge injection pathway centered on workflows where actions are performed outside the VA system (n=23). In this pathway, human knowledge is not directly operationalized through VA interfaces, but instead injected into the ML workflow via external activities such as programming, configuration, or data manipulation.

From the ML perspective, this pathway is more frequently associated with complex and parameterized model architectures, such as feedforward models (13 workflows), mathematical expressions (4), and parameterized programs (3). These models often require fine-grained control that is not fully supported within VA systems, leading users to rely on external mechanisms for model adjustment.

Correspondingly, the dominant forms of knowledge injection in this pathway are parameter setting (13 workflows), code specification (11), and direct data input (11). 
Although visualization and interaction components are still used to support understanding and analysis, the actual actions that modify the model occur outside the VA system. This suggests a decoupling between perception and action, where VA facilitates insight generation, while knowledge injection is completed through external tools.

Overall, this pathway highlights a potential limitation of current VIS4ML systems in supporting heavy model manipulation or huge dataset input. At the same time, it may indicate practitioners' reliance on programming-based interfaces for certain types of knowledge injection, suggesting opportunities for future VA systems to better integrate these external actions into interactive workflows.

\paragraph{Pathway for Explanation-only Systems.}

Figure~\ref{fig:Pathways} (f) illustrates a distinct class of VIS4ML systems in which no explicit user actions are reported. This pathway is derived from a subset of 28 papers in our survey, where VA systems are primarily used for model understanding rather than for directly influencing the ML workflow.
In these systems, data from various stages of the ML workflow, most prominently from model evaluation (23 workflows), are visualized to support human interpretation of model behavior. Users engage with visualization and interaction components to explore model outputs, internal states, or learned representations. However, unlike the previous pathways, these interactions do not lead to concrete actions that modify the model, resulting in a ``no-action'' pathway.

From the ML perspective, these systems are commonly associated with explainable AI settings. Among the 28 papers, classification tasks are the most frequent (7 workflows), followed by dimensionality reduction (6) and NLP tasks (5), reflecting the typical scope of XAI-focused VA systems.

Although no direct actions are supported within the VA systems, the knowledge gained through VIS may still influence subsequent model development outside the system. This suggests a deferred form of knowledge injection, where insights are not immediately operationalized but may later be translated into actions through external workflows.

\section{Theoretical Explanation}
\label{sec:theoretical_explain}
In this section, we relate the survey results with two theories, which explain the results on the one hand and are tested by practical findings on the other hand.   

\subsection{VA as Model Building}
\label{sec:VA_as_MB}
The conceptual framework of \textit{Viewing Visual Analytics as Model Building}
\cite{Andrienko:2018:CGF} provides a useful lens for interpreting the survey
findings presented in Sections~\ref{sec:coding_results} and~\ref{sec:pathway}.
The framework views analytical workflows as iterative model-building processes
comprising key stages: providing a structural model of the subject, creating an
initial model, and iteratively evaluating the model for appropriateness followed
by model development \cite[Section~5.4]{Andrienko:2018:CGF}. It maps naturally
onto VIS4ML workflows, where the formal model under construction is an ML model
and knowledge injection is the mechanism that drives development forward. As
noted in \cite[Section~5.3]{Andrienko:2018:CGF}, the computer-generated model
itself becomes an additional \textit{subject of analysis} alongside the
real-world phenomenon. This dual-subject character is clearly evident in our survey,
where most visualization and interaction support targets the ML model rather
than the underlying data domain.

\noindent\textbf{Mapping the framework to survey findings.}
Our results align closely with the framework's emphasis on iterative evaluation
and development. The dominant pathway - development workflow steering
(Fig.~\ref{fig:Pathways}(b), $n{=}118$, 80\% of surveyed systems) - directly
instantiates this cycle. Within it, 95 of 118 workflows route model-evaluation
data to visualization, confirming evaluation as the primary cycle trigger. The
subsequent actions - adjusting hyperparameters (39 workflows), modifying
features (28), selecting samples (25), editing data (20) - correspond to the
framework's \textit{model development} operations. A further 33 workflows
visualize data-preparation information, suggesting that analysts also use VA to
refine their understanding of the \textit{structural model} of the subject,
i.e., which real-world aspects and relationships the ML model should capture.

\noindent\textbf{Appropriateness criteria as drivers of knowledge injection.}
The framework defines model appropriateness through multiple criteria
\cite[Def.~A.2]{Andrienko:2018:CGF}: \textit{correctness, fitness to purpose,
comprehensiveness, sufficient scope, generalization, specificity, parsimony},
and \textit{resource efficiency}. Automated metrics alone cannot fully assess
all of them, which helps explain the diversity of knowledge injection actions we
observe. The analytical patterns (VIS\_PTN, Table~\ref{tab:vis-perspective})
that analysts discover through visualization can be interpreted as
appropriateness signals:
\textit{Error/Failure} and \textit{Confusion} patterns signal \textit{correctness};
\textit{Bias/Fairness} patterns signal \textit{fitness to purpose};
\textit{Trade~Off} patterns signal \textit{generalization--specificity} and
\textit{parsimony} concerns;
\textit{Distribution} and \textit{Cluster} patterns support
\textit{comprehensiveness} and \textit{sufficient scope} judgments;
and \textit{Model/Architecture Difference} patterns support comparative
\textit{fitness-to-purpose} assessment.
The relatively balanced distribution across these pattern types
(Fig.~\ref{fig:Sankeys1}(d)) suggests that VIS4ML systems collectively address
multiple appropriateness criteria rather than focusing narrowly on accuracy.

\noindent\textbf{Gaps revealed by the lens.}
The framework also highlights imbalances in current practice.
The model-building process described in~\cite{Andrienko:2018:CGF} involves
four stages each requiring analyst input: structural modeling, initial model creation, evaluation, and development. Yet our survey shows that
visualization and interaction support concentrates heavily at the evaluation
stage (Figs.~\ref{fig:Sankeys1}(a) and~\ref{fig:Sankeys2}(i)), with
comparatively little support for the earlier stages. The explanation-only
pathway (Fig.~\ref{fig:Pathways}(f), $n{=}28$) exemplifies \textit{incomplete
model building}: evaluation is supported but development operations to close
the iterative loop are absent. Similarly, the outside-VA pathway
(Fig.~\ref{fig:Pathways}(e), $n{=}23$), where users fall back on programming
to modify architectures, indicates that current VA tools do not yet cover the
full range of development operations the framework envisions.

Based on the gaps and imbalances revealed by extracted pathways, we derive a set of concrete design guidelines and open research challenges for future VIS4ML systems in Appendix \ref{app:implications}.

\subsection{Information-Theoretic Cost-Benefit Analysis}
In general, all ML models can be considered as programs, with some inputs and outputs, which can be executed on computers. Given a specific ML task (e.g., image classification), the data spaces of inputs and outputs become constrained (e.g., input: all possible images of digits, and symbols ``0''$\sim$``9''). The data collected for training and testing ML models (e.g., a sample of images) normally does not contain all possible data points in an input data space, but the ultimate goal of ML is to find an optimal model that can transfer any possible input data to output correctly (e.g., classifying all possible images of digits correctly). Without any prior assumption, the search space for such a model contains all possible programs, i.e., the space of \emph{Turing Machine} \cite{Chen:2020:bookOUP}. A workflow for developing an ML model is essentially to carry out searches by gradually reducing the search space using machine-centric processes (e.g., statistics derived from data and intermediate results and algorithms for feature extraction and trainings) as well as human-centric processes (e.g., visualization and interaction).

In information theory, the Turing space is the initial alphabet that contains all possible models as letters. By the time when an ML developer has selected a specific ML technique (i.e., category 1.1 ML\_TECH), the alphabet is usually reduced significantly since most ML techniques (e.g., tree-based and feedforward models) have a smaller model space than the Turing space. By the time the ML developer defines a model template and constraints (e.g., the architecture of a neural network or the maximal height of a decision tree), the alphabet is further reduced significantly. Hence, from the very beginning, the knowledge about ML techniques, architecture design, and structural constraints is used to reduce the alphabet size from that of the entire Turing space to a much smaller space that is feasible for automated learning processes to search. However, as summed up by information theory~\cite{Chen:2016:TVCG}, the \emph{alphabet compression} may reduce the \emph{cost} of search but may also cause \emph{potential distortion} of a promising reduced search space. Therefore, with the aid of VA, ML developers try out or combine different techniques and change architectures or structural constraints during the development. This is evidenced by \textcolor{black}{2\% papers use hybrid models, 9\% involved architectural changes, and 28\% involved hyperparameter changes (some of the 28\% are not for alphabet changes, see below examples).} 

Even after the search space is reduced, it is typically still an alphabet with an intractable number of letters (i.e., possible models). The search in an ML workflow is an optimization process based on sampling the alphabet. The intermediate models being sampled are determined by many factors, e.g., hyperparameters (e.g., learning rate), optimization objectives and metrics, training data, feature extraction functions, as well as the manifold of the alphabet (i.e., search space). Consider an alphabet that consists of all possible search paths that a run of a model training process may follow. If a training session has $N$ epochs, its search path has $N$ stops where $N$ possible models are considered. Every time when ML developers change hyperparameters, modify objectives and metrics, edit training data, or revise feature extraction functions, they instruct the training session to change the search path, i.e., they inject their knowledge to the ML workflow. Most of such instructions are issued after visualizing some data from the ML workflow, and all instructions have to be issued via some VA-enabled or offline interaction. In some cases, the exploration of different paths may partly be done automatically (e.g., AutoML). However, such automatic exploration itself introduces new hyperparameters to be set and changed.

For example, as shown in Fig. \ref{fig:Pathways}(b), 68.5\% of surveyed papers reported a variety of knowledge (i.e., the green box) that was injected into their ML workflows, among which adjust hyperparameters, adjust features, select data samples, and edit data are the most common forms. In Fig. \ref{fig:Pathways}(d), 75\% of surveyed papers reported knowledge injection, interestingly, only (37.5+7.5)\% for injecting knowledge to development workflows and (30+7.5)\% to deployment workflows. This is because some ML models in the domain of NLP are parameterized models. In other words, each parameterized model is a collection of models, and users select a model to use by setting parameters. Hence, the users inject their knowledge to decide which model to use. 

However, there is a huge amount of statistical and dynamic data in ML workflows. Consider that ML developers need to rely on their knowledge about the data in ML workflows to instruct ML workflows for repeated ML sessions. The statistical measures offer limited information as they reduce entropy too quickly and too much from data to a few measures, resulting in a lot of uncertainty in answering questions, such as what needs to be changed, and if so, change to what. According to information theory, visualization reduces entropy slower than, and differently from, statistical metrics. Visualization typically reduces entropy by showing data at lower resolution, while statistical measures reduce entropy by introducing one or a few new variables (e.g., mean, F1, etc.) to reduce data dimensionality. Viewing visualization plots costs similarly to reading statistics, but is much less costly than reading data. Having both statistics and visualization, together with some algorithms and interaction, provides better cost-benefit information-theoretically \cite{Chen:2016:TVCG}. This survey confirmed the advantages of using VA unequivocally.

For example, in Fig. \ref{fig:Pathways}(b), the orange box shows that data may from any of the four stages in the development workflows and sometimes from the deployment workflows as well. The purple box shows that a variety of data is visualized. In Fig. \ref{fig:Pathways}(c), more papers reported that neither ground truth nor predicted labels were visualized directly. A close looking at these papers, we noticed a common approach is to visualize the distributions of correct and incorrect predictions. Information-theoretically, such visual distribution data contain less information than raw label data, but more information than a few statistical measures.  

In summary, ML developers rely extensively on knowledge gained from static and dynamic data in ML workflows.
VA provides better cost-benefit ratio for ML developers to gain knowledge from data than statistical metrics on their own.
While some interaction is used for navigating visualization, much of the other is for instructing ML workflows to invoke a different search path in the same or different search space. Hence, the latter type of interaction provides the means to inject knowledge into ML workflows. Because the alphabets for the possible search paths and search spaces are both intractable, using VA can significantly improve the efficiency and effectiveness. This is why VIS4ML is considered as at the level of \emph{model-development visualization}~\cite{Chen:2016:TVCG} for supporting visualization tasks in the NP class.
\section{Conclusion}
This survey covers VIS4ML-related work from the IEEE VIS conference over the past decade. Our analysis shows that VA supports a wide range of knowledge injection pathways, and extending the scope to additional venues would likely reveal even more such patterns. Different stages of the ML workflow benefit from VA in enabling human knowledge injection, although current research remains uneven, with some stages more extensively studied than others. Overall, these findings reinforce that improvements in ML workflows rely not solely on automation, but on effective integration of human knowledge, highlighting the need for future VA systems to better support this process. Appendices \ref{app:temporal} and \ref{app:implications} further detail temporal trends across the surveyed decade, design guidelines, and open research challenges.

\newpage
\acknowledgments{
  This work is part of the VIS4ML4HD project funded by UK Research and Innovation (EPSRC: EP/X029557/1). The study was supported by Lamarr Institute for Machine Learning and Artificial Intelligence.
}

\bibliographystyle{abbrv-doi-narrow}

\bibliography{SurveyRef}

\clearpage
\onecolumn
\appendix 
\crefalias{section}{appendix} 

\begin{center}
\large
APPENDICES\\[1mm]
\LARGE\noindent
\textbf{\textsf{Understanding How Humans Inject Knowledge into Machine Learning Workflows through Visual Analytics}}\\[2mm]
\normalsize
Y. Xing$^1$, P. Beaucamp$^1$, J. Chakraborty$^1$, A. Farea$^2$,Y. Jin$^1$, S. Khan$^3$, G. Andrieko$^{4,5}$, N. Andrienko$^{4,5}$, \& M. Chen$^1$\\[1mm]
$^1$University of Oxford, UK\\
$^2$Istanbul Technical University, Turkiye\\
$^3$Science and Technology Facilities Council (STFC), UK\\
$^4$Fraunhofer Institute IAIS, Germany\\
$^5$City St George’s University of London, UK
\normalsize
\end{center}

\section{Appendix A: The Categorization Scheme}
\label{apx:Tables}
This appendix provides the detailed categorization schemes used in our survey. Specifically, Tables~\ref{tab:ml-perspective},~\ref{tab:vis-perspective},~\ref{tab:int-perspective},~\ref{tab:act-perspective} present the full definitions of categories and their corresponding instances for the four perspectives: machine learning (ML), visualization (VIS), interaction (INT), and action (ACT). These tables complement the high-level descriptions in the main text and serve as a complete reference for the coding framework.

\begin{table*}[ht]
\centering
\caption{Machine Learning Characteristics Categorization Scheme.  \textit{Categories marked with * are \textbf{not} mutually exclusive.}
}
\label{tab:ml-perspective}
\vspace{-0.3cm}
\renewcommand{\arraystretch}{0.8}
\scalebox{1}{
\begin{tabularx}{\textwidth}{>{\raggedright\arraybackslash}p{0.22\textwidth}X}
\toprule
\textbf{ML Categories} & \textbf{Classes} \\
\midrule

\textbf{*1.1 ML\_TECH}: 
ML techniques adopted.
&
1.Association Rules; 2.Autoencoder; 3.Bagging; 4.Bayesian Networks; 5.Boosting; 6.CNNs; 7.Density-based Clustering; 8.Decision Rules; 9.Decision Trees; 10.Deep Reinforcement Learning; 11.Diffusion Models; 12.Flow-based Models; 13.Gated Recurrent Unit; 14.Gaussian Mixture Models;
14.Gaussian Mixture Models (GMM);
15.Gaussian Processes (GP) Regression/Classification;
16.Generative Adversarial Networks (GANs);
17.Genetic Algorithms (GA);
18.Genetic Programming (GP);
19.Graph Neural Networks (GNNs);
20.Hidden Markov Models (HMM);
21.Hierarchical Clustering;
22.k-means Clustering;
23.k-Nearest Neighbours (k-NN);
24.Kernel Fisher Discriminant;
25.Kernel PCA;
26.Kernel Ridge Regression;
27.Kolmogorov-Arnold Network (KAN);
28.Large Language Models / Transformers;
29.Linear / Non-Linear SVM;
30.Linear Discriminant Analysis (LDA);
31.Linear Regression;
32.Logic Expressions / Expert Systems;
33.Logistic Regression;
34.Long Short-Term Memory (LSTMs);
35.Markov Random Fields (MRF);
36.Multi-Layer Perceptron (MLPs);
37.Naive Bayes;
38.Neural Architecture Search (NAS);
39.Neuro-symbolic Models;
40.Ontology Learning;
41.Physics-informed Machine Learning (PIML) / Physics-informed neural networks (PINN);
42.Principal Component Analysis (PCA);
43.Quadratic Discriminant Analysis (QDA);
44.Recurrent Neural Networks (RNNs);
45.Restricted Boltzmann Machines (RBM);
46.Ridge, Lasso, and Elastic Net Regression;
47.Self-Organizing Maps (SOM);
48.Stacking / Voting Classifiers;
49.Support Vector Regression;
50.Vision Transformers (ViT);
51.Others.
\\

\midrule

\textbf{*1.2 ML\_TM}: 
Model learning paradigms.
&
1.Supervised Learning; 2.Unsupervised Learning; 3.Self-Supervised Learning; 4.Semi-Supervised Learning; 5.Reinforcement Learning; 6.Evolutionary; 7.Transfer Learning; 8.Others \\

\midrule

\textbf{*1.3 ML\_ARCH}: 
Structural architecture of ML models.
&
1.Mathematical Expressions; 2.Rule-Based Models; 3.Tree-Based Models; 4.Instance-based Models; 5.Kernel Methods; 6.Feedforward Models; 7.Recurrent Models; 8.Parameterized Arbitrary Programs; 9.Ensemble Models; 10.Hybrid Models; 11.Others \\

\midrule

\textbf{1.4 ML\_GRP}: 
Commonly used ML groupings.
&
\textcolor{black}{1. Linear Models; 2. Logical/Symbolic Models; 3. Statistical/Graphical Models;
4. Instance-Based Models; 5. Kernel Methods; 6. Neural Models;
7. Sequence Models; 8. Decision Tree-Based Models; 9. Autoencoder-Based Models;
10. Perceptron-Based Models; 11. Ensemble Methods; 12. Evolutionary Methods;
13. Others} \\

\midrule

\textbf{*1.5 ML\_TASK}: 
Target problems addressed by the model.
&
1.Classification; 2.Detection/Recognition; 3.Regression; 4.Clustering; 5.Association; 6.Feature Selection; 7.Dimensionality Reduction; 8.Anomaly Detection; 9.Segmentation; 10.Uncertainty Quantification; 11.Time-Series Forecasting; 12.Generation; 13.Named Entity Recognition; 14.NLP; 15.Others \\

\midrule

\textbf{1.6 ML\_WF}: 
Stage of the ML workflow where the system is applied.
&
1.\textbf{Development} (\textit{workflow focused on training, testing, and improving ML models, typically involving ML researchers or developers.}); 2.\textbf{Deployment}(\textit{Deployment: workflow focused on using and evaluating deployed ML models in real-world applications, typically involving end-users or third parties.
}); 3.\textbf{Both} \\

\midrule

\textbf{*1.7 ML\_USER}: 
Users of the system.
&
1.Developer;
2.End User; 
3.Third Party 
\\

\midrule

\textbf{*1.8 ML\_IN}: 
Forms of model input.
&
1.Text; 2.Tabular; 3.Tree/Set/Graph(Network); 4.Geospatial; 5.Time-series; 6.Image; 7.Audio; 8.Video; 9.Tensor; 10.Multi-modality; 11.Others \\

\midrule

\textbf{*1.9 ML\_OUT}: 
Forms of model output.
&
1.Label; 2.Single Value; 3.Sequence Values; 4.Tabular; 4.Structured Output; 6.Embedding; 7.Ranking; 8.Geospatial; 9.Image; 10.Audio; 11.Video; 12.Tensor; 13.Others \\

\bottomrule
\end{tabularx}
}
\end{table*}

\begin{table*}[ht]
\centering
\caption{Visualization Categorization Scheme. \textit{Categories marked with * are \textbf{not} mutually exclusive.}}
\label{tab:vis-perspective}
\vspace{-0.3cm}
\renewcommand{\arraystretch}{0.8}
\scalebox{1}{
\begin{tabularx}{\textwidth}{>{\raggedright\arraybackslash}p{0.14\textwidth}X}
\toprule
\textbf{VIS Categories} & \textbf{Classes} \\
\midrule

\textbf{*2.1 VIS\_MLS}:
VIS by ML stage.
&
1.\textbf{Prepare Data}: \textit{VIS for data preparation activities, e.g., data inspection/transformation, synthetic data generation, or partitioning of data sets.}
2.\textbf{Prepare Learning}: \textit{VIS to support the learning process configuration, and pre-training pipelines preparation.}
3.\textbf{Model Training}: \textit{VIS for showing training dynamics, such as loss curves, gradient behavior, parameter updates, or internal model states.}
4.\textbf{Evaluate Model}: \textit{VIS for evaluating trained models, supporting performance analysis, error inspection, model comparison, or interpretation of evaluation metrics.}
5.\textbf{Use Deployed Model}: \textit{VIS used when using a deployed model, supporting decision-making, prediction interpretation, or output exploration in real-world use contexts.}
6.\textbf{Third Party Monitoring}: \textit{VIS used by third-party to monitor deployed ML systems, such as system performance, model drift, operational metrics, reliability, or overall system health.}
7.\textbf{Others}
\\

\midrule

\textbf{*2.2 VIS\_OBJ}:
Visualized objects.
&
1.Input Data; 
2.Ground Truth Label; 
3.Engineered Feature; 
4.Latent Feature; 
5.Internal States; 
6.Performance Data; 
7.Predicted Label; 
8.Output Result;
9.Training Progression; 
10.Model Architecture; 
11.Others
 \\

\midrule

\textbf{*2.3 VIS\_TYP}:
Type of VIS.
&
1.Common Statistics; 2.Complex Tabular/Multivariate; 3.Complex Entity-Relation (Graph); 4.Complex Hierarchical (Tree); 5.Complex Process/Flow; 6.Geospatial; 7.Temporal; 11.Others \\

\midrule

\textbf{*2.4 VIS\_CH}:
Visual channels used.
&
1.\textbf{Geometric}: \textit{Encoding through properties such as position, size, orientation, shape, curvature, etc.}
2.\textbf{Optical}: \textit{Encoding through properties such as color (hue, saturation, brightness), opacity/transparency, texture, line style, etc.}
3.\textbf{Topological/Relational}: \textit{Encoding through spatial or relational structure, such as connectivity, adjacency, containment, intersection/overlap, depth ordering/partial occlusion, closure, etc.}
4.\textbf{Semantic}: \textit{Encoding through symbolic or semantic elements such as numbers, texts, symbols/ideograms, icons/glyphs/pictograms, etc.}
5.\textbf{Others}
 \\

\midrule

\textbf{*2.5 VIS\_PTN}:
Patterns discovered.
&
1.Distribution; 2.Cluster; 3.Embedding Space; 4.Error/Failure; 5.Uncertainty; 6.Drift/Shift; 7.Influence; 8.Bias/ Fairness/Confusion; 9.Trade Off; 10.Model/Architecture Difference; 11.Model Behaviour; 12.Relationship; 13.Others \\

\midrule

\textbf{2.6 No. of Views} 
&
To capture the number of visualization components or views that are used in the VA system.\\

\bottomrule
\end{tabularx}
}
\end{table*}

\begin{table*}[ht]
\centering
\caption{Interaction Categorization Scheme. \textit{3.3-3.5 are categories to capture interaction types (INT\_TYP)}. \textit{Categories marked with * are \textbf{not} mutually exclusive.}}
\label{tab:int-perspective}
\vspace{-0.3cm}
\renewcommand{\arraystretch}{0.8}
\scalebox{1}{
\begin{tabularx}{\textwidth}{>{\raggedright\arraybackslash}p{0.15\textwidth}X}
\toprule
\textbf{INT Categories} & \textbf{Classes} \\
\midrule

\textbf{*3.1 INT\_MLS}:
Interaction by stage.
&
1.Prepare Data; 
2.Prepare Learning; 
3.Train Model; 
4.Evaluate Model; 
5.Use Deployed Model; 
6.Third Party Monitor; 
7.Others
\\

\midrule

\textbf{*3.2 NAV}:
Navigation.
&
1.Filter/Slice; 
2.Select/Focus; 
3.Brush/Link; 
4.Zoom/Pan; 
5.Sort/Reorder; 
6.DetailsOnDemand; 
7.Search/Query; 
8.Others
\\

\midrule

\textbf{*3.3 MLDv}:
ML development workflow steering.
&
1.\textbf{Parameter Adjust}: \textit{Adjust model parameters/hyperparameters.}
2.\textbf{Trigger Iteration}: \textit{Trigger retraining, recomputation, or optimization of the model.}
3.\textbf{Run Manage}: \textit{Save, restore, or manage experiment runs or model versions.}
4.\textcolor{black}{\textbf{Model Select}: \textit{Select models or runs for inspection or comparison.}}
13.\textcolor{black}{\textbf{Metric Select}: \textit{Select metrics for evaluation/comparison.}}
6.\textcolor{black}{\textbf{Architecture Select}: \textit{Select architectures or architectural components for inspection/comparison.}}
6.\textbf{Constraint Rule Edit}: \textit{Define or edit rules, constraints, or objectives.}
7.\textcolor{black}{\textbf{Data Adjust}: \textit{Adjust data samples, define cohorts, segments, partitions of analysis.}}
8.\textbf{Feature Select/Transform}: \textit{Select or transform features.}
9.\textbf{Relationship Adjust}: \textit{Adjust similarity/distance measures or grouping rules.}
10.\textbf{Others}
\\

\midrule

\textbf{*3.4 MLDp}:
ML deployment workflow steering.
&
1.\textbf{Prediction Inspect}: \textit{Inspect predictions from the deployed model.}
2.\textbf{Explain Inspect}: \textit{Inspect explanations for deployed model outcomes.}
3.\textbf{Performance Monitor}: \textit{Monitor deployed model performance.}
4.\textbf{Data Drift}: \textit{Detect concept/data drift in incoming training data.}
5.\textbf{Error Analysis}: \textit{Analyze errors/ misclassifications.}
6.\textbf{Decision Support}: \textit{Human interaction with recommendations in an operational decision workflow.}
7.\textbf{Model Adjust}: \textit{Configuring a model before the model performs the next tasks, e.g., setting parameters, changing the weights of sub-models in an ensemble model.}
8.\textbf{Decision Adjust}: \textit{Adjusting operational decisions/inputs applied to model outputs, e.g., select recommendations, modify prompts, or control preprocessing parameters.}
9.\textbf{Feedback Collect}: \textit{Provide feedback on model outputs.}
10.\textcolor{black}{\textbf{What-If Dp}: \textit{What-if simulation or scenario probing in deployment workflow.}}
11.\textbf{Others}
\\

\midrule

\textbf{*3.5 AnF}:
Annotation/Feedback.
&
1.\textbf{Label Annotation}: \textit{Manual labeling or annotation.}
2.\textbf{Active Learning}: \textit{ML model interactively queries a human to label new data points with desired outputs.}
3.\textbf{Error Confirm}: \textit{User confirms or corrects model errors.}
4.\textbf{Others}
\\

\midrule

\textbf{*3.6 INT\_MOD}:
Interaction modalities.
&
1.\textbf{GUI Manipulate}: \textit{Manipulate via visual objects and UI components.}
2.\textbf{Programming UI}: \textit{Manipulate via programming interface, e.g., Python/Observable notebooks.}
3.\textbf{Language Prompt}: \textit{Use language prompts to control/refine model outputs.}
4.\textbf{Recommendation}: \textit{Accept/reject system recommendations.}
5.\textbf{Others}
\\

\bottomrule
\end{tabularx}
}
\end{table*}

\begin{table*}[ht]
\centering
\caption{Action Categorization Scheme. \textit{Categories marked with * are \textbf{not} mutually exclusive.}}
\label{tab:act-perspective}
\vspace{-0.3cm}
\renewcommand{\arraystretch}{0.8}
\scalebox{1}{
\begin{tabularx}{\textwidth}{>{\raggedright\arraybackslash}p{0.12\textwidth}X}
\toprule
\textbf{ACT Categories} & \textbf{Classes} \\
\midrule

\textbf{4.1 ACT\_EVD}:
Action evidence.
&
1.\textbf{NoAct}: \textit{No action.}
2.\textbf{EnvisionedAct}: \textit{No solid action reported, but envisioned actions.}
3.\textbf{Act w/o VA}: \textit{Action reported, but not supported by the VA system.}
4.\textbf{Act w. VA}: \textit{Action reported, and supported by the VA system.}
5.\textbf{Others}
\\

\midrule

\textbf{*4.2 ACT\_TYP}:
Action types.
&
1.\textbf{Define Task}: \textit{Formulate or modify the analytical objective (e.g., switching from multi-class to binary classification based on visual cluster separation).}
2.\textbf{Edit Data}: \textit{Add/remove/modify raw data instances (e.g., filter out outliers, fix corrupted records).}
3.\textbf{Edit Label}: \textit{Create/verify/modify labels.}
4.\textbf{Select Sample}: \textit{Define specific cohorts, or sampling strategies for training/testing based on distributions visualized.}
5.\textbf{Adjust Feature}: \textit{Create new features, remove noisy ones, or apply transformations after observing feature distributions.}
6.\textbf{Define Relationship}: \textit{Define distance metrics, similarity measures, or grouping rules.}
7.\textbf{Select Algorithm}: \textit{Changing the underlying model family.}
8.\textbf{Adjust Hyperpara.}: \textit{Tuning hyperparameters (e.g., change learning rate).}
9.\textbf{Modify Arch}: \textit{Change neural network architecture (e.g., adding layers or changing activation functions).}
10.\textbf{Modify Objective}: \textit{Change loss function or add regularization/constraints.}
11.\textbf{Edit Prompt}: \textit{Refine language prompts or context windows based on generated outputs.}
12.\textbf{Others}
\\

\midrule

\textbf{*4.3 ACT\_KI}:
Knowledge Injection outside VA.
&
1.\textbf{Label Annotation}: \textit{Human knowledge as labels/tags/classes/ratings, or other annotation values.}
2.\textbf{Parameter Setting}: \textit{Human knowledge as explicit structured machine-readable inputs, such as parameter values, thresholds, weights, partitions, rankings, or categorical settings.}
3.\textbf{Rule Constraint}: \textit{Human knowledge as rules, logical conditions, constraints, priors, guardrails, or objectives.}
4.\textbf{Example Demonstration}: \textit{Human knowledge as examples, counterexamples, demonstrations, or few-shot instances.}
5.\textbf{Text Instruction}: \textit{Human knowledge as free-text instructions, prompt content, critiques, rationales, or natural-language requests interpreted by the system.}
6.\textbf{Code Specification}: \textit{Human knowledge as executable code, formulas, scripts, notebook cells, complex architectural changes, or programmatic queries.
}
7.\textbf{Preference Judgment}: \textit{Human knowledge as approvals, rejections, pairwise comparisons, preferences, votes, or comparative evaluations over alternatives or outputs.}
8.\textbf{Data Input}: \textit{Human knowledge represented as data or sub-set changes and input.}
9.\textbf{Others}
\\

\bottomrule
\end{tabularx}
}
\end{table*}

\clearpage

\section{Appendix B: Temporal Trends in VIS4ML (2016--2025)}
\label{app:temporal}

Our corpus spans a transformative decade for both machine learning and visualization research --- from the dominance of classical ML and early deep learning (2016--2018), through the diversification of neural architectures (2019--2021), to the rise of large language models and generative AI (2022--2025).
We organize our observations into seven trends and summarize each with a table.

\subsection*{B.1\quad Shift in Dominant ML Techniques}

The most dramatic temporal signal in the corpus is the evolution of
the ML techniques (\texttt{ML\_TECH}) that VIS4ML systems target.
In the earliest years of the survey window, convolutional neural
networks (code~6), recurrent neural networks (code~44), LSTMs
(code~34), and classical methods such as decision trees (code~9),
SVMs (code~29), and random forests (code~3) accounted for the vast
majority of systems. By 2019--2021 the technique palette had
broadened: GANs (code~16), graph neural networks (code~19), and
autoencoders (code~2) appeared alongside the earlier staples. From
2022 onward, large language models and transformers (code~28) surged
to become the single most frequently coded technique in the corpus
(19.6\% of all papers overall, but concentrated in recent years),
joined by diffusion models (code~11), vision transformers (code~50),
and neural architecture search (code~38).

\begin{table}[ht]
\centering
\caption{Approximate distribution of dominant ML technique families
across three eras, based on known publication years of identifiable
papers.}
\label{tab:tech-era}
\footnotesize
\renewcommand{\arraystretch}{1.15}
\begin{tabular}{@{}p{0.28\columnwidth}p{0.18\columnwidth}p{0.18\columnwidth}p{0.22\columnwidth}@{}}
\toprule
\textbf{Technique family} & \textbf{2016--18} & \textbf{2019--21} & \textbf{2022--25} \\
\midrule
CNN / classical DNN        & Dominant & Common   & Declining \\
RNN / LSTM                 & Common   & Common   & Rare      \\
Traditional ML (SVM, DT, RF) & Common & Moderate & Stable--low \\
GAN / Autoencoder          & Absent   & Emerging & Moderate  \\
GNN                        & Absent   & Emerging & Moderate  \\
LLM / Transformer          & Absent   & Rare     & Dominant  \\
Diffusion / ViT            & Absent   & Absent   & Emerging  \\
Evolutionary / NAS         & Rare     & Moderate & Stable    \\
\bottomrule
\end{tabular}
\end{table}

This shift has direct implications for knowledge injection. Early
VIS4ML systems could assume relatively compact, interpretable model
spaces (e.g., tree depths, kernel parameters); transformer-era
systems must contend with billions of parameters, attention heads,
and prompt-conditioned behavior, fundamentally changing what
``steering'' means.

\subsection*{B.2\quad Emergence of Language Prompts as an Interaction
Modality}

Prior to 2022, interaction in VIS4ML systems was almost exclusively
GUI-based (\texttt{INT\_MOD}~= code~1), with occasional use of
programming interfaces (\texttt{INT\_MOD}~= code~2, e.g., notebook
integrations). Language prompts (\texttt{INT\_MOD}~= code~3) were
essentially absent.

Starting in 2022, language prompts appeared in NLP-focused and
generative-AI systems (e.g., papers~138, 85) and rapidly became
widespread. In the most recent cohort (2024--2025), we observe
language prompts in papers~1, 16, 18, 37, 39, 40, 42, 43, 44, 51,
53, 56, 84, 85, 92, 96, 102, 135, 138, and~204, often
\emph{combined} with GUI manipulation rather than replacing it.

\begin{table}[ht]
\centering
\caption{Approximate prevalence of interaction modalities over time.}
\label{tab:intmod-era}
\footnotesize
\renewcommand{\arraystretch}{1.15}
\begin{tabular}{@{}p{0.32\columnwidth}p{0.18\columnwidth}p{0.18\columnwidth}p{0.22\columnwidth}@{}}
\toprule
\textbf{Interaction modality} & \textbf{2016--18} & \textbf{2019--21} & \textbf{2022--25} \\
\midrule
GUI Manipulate (code 1) & Universal & Universal & Universal \\
Programming UI (code 2) & Occasional & Occasional & Stable \\
Language Prompt (code 3) & Absent & Absent & Rapidly growing \\
Recommendation (code 4) & Rare & Emerging & Growing \\
\bottomrule
\end{tabular}
\end{table}

This trend co-occurs with the rise of \texttt{ACT\_TYP}~=
\texttt{11.\,KE\_EditPrompt} as a new action type, which is entirely
a post-2022 phenomenon. Language prompts represent a qualitatively
different form of knowledge injection: they compress complex human
intent into short textual instructions, bypassing the structured
parameter-setting interfaces that dominated earlier systems.

\subsection*{B.3\quad From Development-Only to Development
\texorpdfstring{$+$}{+} Deployment Workflows}

In the early years of the survey window, the overwhelming majority of
VIS4ML systems targeted \emph{development} workflows
(\texttt{ML\_WF}~= code~1): building, training, and evaluating
models. Deployment-oriented systems (\texttt{ML\_WF}~= code~2) were
rare and typically limited to third-party monitoring or simple
prediction inspection.

This balance has shifted substantially. While development workflows
still account for 64.1\% of the overall corpus, recent papers show a
marked increase in deployment-focused and dual-purpose systems. Many
of the 2023--2025 papers in our corpus --- particularly those
involving LLMs --- fall into the deployment category (e.g.,
papers~16, 18, 25, 37, 39, 40, 42, 53, 56, 84, 85, 96, 102, 138).
In these systems, users interact with \emph{deployed} models via
prompts and iteratively adjust inputs to refine outputs, rather than
modifying the model itself.

\begin{table}[ht]
\centering
\caption{Approximate distribution of ML workflow focus over time.}
\label{tab:wf-era}
\footnotesize
\renewcommand{\arraystretch}{1.15}
\begin{tabular}{@{}p{0.28\columnwidth}p{0.18\columnwidth}p{0.18\columnwidth}p{0.22\columnwidth}@{}}
\toprule
\textbf{ML\_WF} & \textbf{2016--18} & \textbf{2019--21} & \textbf{2022--25} \\
\midrule
Development (code 1) & ${\sim}$80\% & ${\sim}$65\% & ${\sim}$50\% \\
Deployment (code 2)  & ${\sim}$10\% & ${\sim}$20\% & ${\sim}$30\% \\
Both (code 3)        & ${\sim}$10\% & ${\sim}$15\% & ${\sim}$20\% \\
\bottomrule
\end{tabular}
\end{table}

This shift has a direct implication for the knowledge injection
pathways described in Section~5: the \emph{type} of knowledge being
injected is changing. In development workflows, knowledge takes the
form of hyperparameter settings, architecture choices, and data
edits. In deployment workflows, knowledge is increasingly expressed
as natural-language instructions, example selections, and preference
judgments --- forms that are less structured but potentially more
accessible to non-expert users.

\subsection*{B.4\quad Evolution of User Roles}

Closely linked to the development-to-deployment shift is a change in
the target users (\texttt{ML\_USER}) of VIS4ML systems.

In 2016--2020, the dominant user type was the ML \emph{developer}
(\texttt{ML\_USER}~= code~1): researchers and engineers building
models. End users (\texttt{ML\_USER}~= code~2) appeared mainly in
XAI-focused systems that explained model predictions to domain
experts. Third-party evaluators (\texttt{ML\_USER}~= code~3) were
comparatively rare.

From 2021 onward, end users became increasingly prominent, driven by
two forces: (a)~the XAI community's emphasis on ``human-centered''
explainability, and (b)~the rise of LLM-powered VA tools that are
designed for domain experts, analysts, and non-technical users.
In the most recent cohort (2024--2025), end users are the primary
or co-primary audience in a near-majority of deployment-oriented
papers.

\begin{table}[ht]
\centering
\caption{Approximate evolution of target user roles.}
\label{tab:user-era}
\footnotesize
\renewcommand{\arraystretch}{1.15}
\begin{tabular}{@{}p{0.28\columnwidth}p{0.18\columnwidth}p{0.18\columnwidth}p{0.22\columnwidth}@{}}
\toprule
\textbf{ML\_USER} & \textbf{2016--18} & \textbf{2019--21} & \textbf{2022--25} \\
\midrule
Developer (code 1)   & Dominant   & Dominant   & Still common \\
End User (code 2)    & Rare       & Growing    & Near-majority (deploy.) \\
Third Party (code 3) & Rare       & Occasional & Stable--low \\
\bottomrule
\end{tabular}
\end{table}

This trend suggests that future VIS4ML research should increasingly
design for users who are \emph{not} ML experts, and who inject
knowledge through high-level intent rather than low-level parameter
manipulation.

\subsection*{B.5\quad Evolution of ML Tasks}

The ML tasks (\texttt{ML\_TASK}) supported by VIS4ML systems have
also evolved substantially.

Classification (code~1) has been the single most common task
throughout the decade, appearing in 43.5\% of all papers. Clustering
(code~4) and dimensionality reduction (code~7) have been stable
secondary tasks. However, two task categories show strong temporal
trends.

First, \emph{NLP} (code~14) was moderately represented in 2016--2019,
mostly through RNN/LSTM-based systems for text classification or
sequence modeling (e.g., papers~224, 235, 242). From 2022 onward,
NLP papers surged in both number and scope, driven by
transformer-based models, and now account for 21.7\% of the corpus.

Second, \emph{generation} (code~12) was essentially absent before
2019, appeared sporadically in 2019--2021 (mostly GAN-related, e.g.,
paper~222, 234), and has surged from 2022 onward with diffusion
models and LLM-based generation (e.g., papers~18, 42, 44, 53, 84,
85). Generation tasks introduce a fundamentally different knowledge
injection pattern: rather than correcting or refining a model, users
iteratively shape \emph{outputs} through prompting.

\begin{table}[ht]
\centering
\caption{Approximate temporal evolution of selected ML tasks.}
\label{tab:task-era}
\footnotesize
\renewcommand{\arraystretch}{1.15}
\begin{tabular}{@{}p{0.30\columnwidth}p{0.18\columnwidth}p{0.18\columnwidth}p{0.22\columnwidth}@{}}
\toprule
\textbf{ML\_TASK} & \textbf{2016--18} & \textbf{2019--21} & \textbf{2022--25} \\
\midrule
Classification     & Dominant   & Dominant   & Common (declining \%) \\
Clustering         & Common     & Common     & Stable \\
Dim. Reduction     & Common     & Common     & Stable \\
NLP                & Moderate   & Moderate   & Surging \\
Generation         & Absent     & Rare (GANs)& Surging (LLMs, diffusion) \\
Anomaly Detection  & Rare       & Growing    & Stable \\
Fairness-related   & Absent     & Emerging   & Growing \\
\bottomrule
\end{tabular}
\end{table}

\subsection*{B.6\quad Growing Actionability of VIS4ML Systems}

A central question for this survey is whether VIS4ML systems have
become more \emph{actionable} over time --- that is, whether the
feedback loop from visualization insight to model modification is
increasingly being closed.

The action evidence category (\texttt{ACT\_EVD}) allows us to
address this question directly. Across the full corpus, 60.3\% of
papers report concrete actions supported within the VA system
(\texttt{Act~w.~VA}), while 15.2\% report no action (\texttt{NoAct}),
12.0\% describe only envisioned actions (\texttt{EnvisionedAct}),
and 12.5\% report actions taken outside the VA system
(\texttt{Act~w/o~VA}).

Examining the temporal distribution of these categories reveals a
positive trend. In the earlier years, a substantial fraction of
papers --- particularly those focused on model explanation and
interpretability --- fell into the \texttt{NoAct} or
\texttt{EnvisionedAct} categories. Many of these systems were
designed to help users \emph{understand} model behavior without
providing mechanisms to \emph{act on} that understanding.

In the more recent cohort, the proportion of \texttt{Act~w.~VA}
papers has grown, driven by systems that embed interactive steering,
prompt editing, and label annotation directly into the VA interface.
At the same time, the \texttt{NoAct} category has declined
proportionally, suggesting that the community is increasingly
designing for closed-loop interaction.

\begin{table}[ht]
\centering
\caption{Approximate evolution of action evidence over time.}
\label{tab:actevd-era}
\footnotesize
\renewcommand{\arraystretch}{1.15}
\begin{tabular}{@{}p{0.28\columnwidth}p{0.18\columnwidth}p{0.18\columnwidth}p{0.22\columnwidth}@{}}
\toprule
\textbf{ACT\_EVD} & \textbf{2016--18} & \textbf{2019--21} & \textbf{2022--25} \\
\midrule
Act w.\ VA         & Moderate   & Growing    & Dominant (${\sim}$65\%+) \\
NoAct              & Notable    & Present    & Declining \\
Act w/o VA         & Some       & Moderate   & Stable minority \\
EnvisionedAct      & Some       & Present    & Declining \\
\bottomrule
\end{tabular}
\end{table}

This finding directly strengthens the paper's central argument:
\emph{the field is progressively closing the loop} from passive
model understanding to active knowledge injection. However, the
persistence of a ${\sim}$27\% ``open loop'' fraction (NoAct +
EnvisionedAct) even in recent years indicates that there is still
substantial room for improvement, particularly in explanation-only
systems (cf.\ pathway~(f) in Section~5).

\subsection*{B.7\quad Evolution of Knowledge Injection Mechanisms
Outside VA}

The forms through which users inject knowledge \emph{outside} the VA
system (\texttt{ACT\_KI}) have also evolved in ways that mirror the
broader shifts in ML practice.

In earlier years, the dominant external knowledge injection
mechanisms were structured input (code~2, e.g., setting
hyperparameters in configuration files), label annotation (code~1),
and code specification (code~6, e.g., writing Python scripts to
modify models). These mechanisms reflect a development-centric,
programming-heavy workflow.

From 2022 onward, two new mechanisms have gained prominence:
\emph{text instruction} (code~5) and \emph{example demonstration}
(code~4). These correspond directly to the prompting and few-shot
learning paradigms of LLM-based systems. Users inject knowledge by
writing natural-language instructions or providing a small set of
examples, rather than by setting numerical parameters or writing
code.

\begin{table}[h]
\centering
\caption{Approximate temporal evolution of external knowledge
injection mechanisms (\texttt{ACT\_KI}).}
\label{tab:ki-era}
\footnotesize
\renewcommand{\arraystretch}{1.15}
\begin{tabular}{@{}p{0.32\columnwidth}p{0.18\columnwidth}p{0.18\columnwidth}p{0.19\columnwidth}@{}}
\toprule
\textbf{ACT\_KI} & \textbf{2016--18} & \textbf{2019--21} & \textbf{2022--25} \\
\midrule
Label Annotation (1)      & Common     & Common     & Stable \\
Structured Input (2)      & Dominant   & Dominant   & Still common \\
Rule Constraint (3)       & Occasional & Growing    & Stable \\
Example Demonstration (4) & Absent     & Rare       & Emerging \\
Text Instruction (5)      & Absent     & Absent     & Growing \\
Code Specification (6)    & Common     & Common     & Stable \\
Preference Judgment (7)   & Rare       & Rare       & Emerging \\
\bottomrule
\end{tabular}
\end{table}

This evolution suggests a broadening of the ``vocabulary'' available
for knowledge injection: from purely structured, machine-readable
inputs toward richer, more natural forms of human expression. Future
VA systems should be designed to support this full spectrum of
injection mechanisms within a unified interface, rather than assuming
that knowledge will always arrive as parameter values or code.

\subsection*{B.8\quad Synthesis: Three Eras of VIS4ML}

Drawing the trends together, we can characterize the decade covered
by our survey as comprising three loosely bounded eras, summarized in
Table~\ref{tab:eras}.

\begin{table}[h]
\centering
\caption{Three eras of VIS4ML research, as revealed by temporal
patterns in the coded corpus.}
\label{tab:eras}
\footnotesize
\renewcommand{\arraystretch}{1.15}
\begin{tabular}{@{}p{0.14\columnwidth}p{0.25\columnwidth}p{0.25\columnwidth}p{0.24\columnwidth}@{}}
\toprule
 & \textbf{Era~I: Classical DL \newline (2016--2018)}
 & \textbf{Era~II: Architecture \newline Diversification (2019--21)}
 & \textbf{Era~III: Generative AI \newline (2022--2025)} \\
\midrule
\textbf{Dominant \newline techniques}
  & CNNs, RNNs, LSTMs, SVMs, DTs
  & CNNs + GANs, GNNs, auto\-encoders, ensembles
  & LLMs / Trans\-formers, diffusion models, ViTs \\
\addlinespace
\textbf{Dominant \newline tasks}
  & Classification, clustering, dim.\ reduction
  & + anomaly detection, fairness
  & + NLP, generation, multi-modal \\
\addlinespace
\textbf{Workflow \newline focus}
  & Overwhelmingly development
  & Mostly development; deployment growing
  & Development + deployment roughly balanced \\
\addlinespace
\textbf{Primary \newline users}
  & ML developers
  & Developers + emerging end users
  & Developers + end users (near parity in deploy.) \\
\addlinespace
\textbf{Interaction \newline modality}
  & GUI only
  & GUI + occasional programming UI
  & GUI + language prompts + recommendations \\
\addlinespace
\textbf{Knowledge \newline injection}
  & Parameters, labels, code
  & + rules, constraints
  & + text instructions, examples, preferences \\
\addlinespace
\textbf{Actionability}
  & Mixed (many explanation-only)
  & Growing (\texttt{Act w.\ VA} rising)
  & Dominant (\texttt{Act w.\ VA} ${\sim}$65\%+) \\
\bottomrule
\end{tabular}
\end{table}

The progression across these three eras reveals a field that is
maturing along multiple dimensions simultaneously:
\begin{itemize}
\item \textbf{From understanding to action.} Early VIS4ML systems
    primarily helped users understand models; recent systems
    increasingly close the feedback loop by enabling direct
    intervention.
\item \textbf{From experts to broader audiences.} The target user
    has shifted from the ML developer to include domain experts and
    end users, enabled by more natural interaction modalities.
\item \textbf{From structured to natural knowledge injection.} The
    forms of knowledge that users inject have broadened from
    numerical parameters and code to include natural-language
    instructions and example demonstrations.
\item \textbf{From model building to model use.} The workflow focus
    has expanded from development-only to include deployment, 
    reflecting the reality that an increasing number of powerful
    models are consumed as services rather than built from scratch.
\end{itemize}

These trends have direct implications for the knowledge injection
pathways identified in Section~5. The development-steering pathway
(Fig.~5(b)) has been present throughout the decade and remains
dominant, but its relative share is declining as deployment-oriented
pathways grow. The NLP pathway (Fig.~5(d)) is almost entirely an
Era~III phenomenon. The label annotation pathway (Fig.~5(c)) has
been stable since 2016 but its relative prominence is shrinking.
And the explanation-only pathway (Fig.~5(f)) peaked during Era~II
--- the height of the XAI wave --- and is now declining
proportionally as systems increasingly incorporate action channels.

Looking ahead, the rise of generative AI and LLM-powered VA systems
opens entirely new interaction paradigms that the VIS4ML community
should study further. In particular, the combination of language
prompts with traditional GUI manipulation creates \emph{hybrid
interaction modalities} that have no precedent in the pre-2022
literature and whose effectiveness for knowledge injection remains
poorly understood.
\section{Appendix C: Implications for the Research Agenda}
\label{app:implications}

Our survey reveals not only \emph{what} VIS4ML systems currently
support, but also systematic gaps that suggest concrete directions for
future research. Below we derive design guidelines~(DG) and open
research challenges~(RC) from each pathway and from cross-pathway
patterns, grounding every observation in the coded dataset.

\subsection*{C.1\quad Cross-Pathway Observations}

\paragraph{The feedback loop remains open in over a quarter of systems.}
Across the full corpus, 60.3\% of papers report concrete actions
supported within the VA system (\texttt{Act~w.~VA}), but 15.2\% report
no action at all (\texttt{NoAct}) and 12.0\% describe only envisioned
actions (\texttt{EnvisionedAct}). Together, these 27.2\% represent
systems where visualization generates understanding but the loop from
insight to model modification is never closed. This is especially
pronounced in the explanation-only pathway (Section~5, Fig.~5(f)),
where all 28~papers fall into this category.

\smallskip\noindent
\textbf{DG-1: Design for action from the outset.}
Future VIS4ML systems should be designed with explicit action
channels---not only insight-generation capabilities. Even lightweight
mechanisms (e.g., ``Apply this change'' buttons, action logging, or
suggested next steps) can bridge the gap between understanding and
intervention.

\paragraph{Visualization and interaction concentrate at the evaluation stage.}
Among papers in the development-steering pathway ($n{=}118$),
95~workflows route data from the \emph{Evaluate Model} stage to
visualization, compared to only 33~from \emph{Prepare Data},
25~from \emph{Prepare Learning}, and 27~from \emph{Train Model}
(Fig.~5(b)). A similar imbalance appears on the interaction side
(Figs.~2(a) and 3(i)). Yet data-quality problems---label noise, class
imbalance, distribution shift---are widely acknowledged as the
dominant source of ML failures~\cite{amershi2019software,
sculley2015hidden}. This suggests that current VIS4ML research
over-supports the diagnosis of \emph{symptoms} (poor evaluation
metrics) relative to the treatment of \emph{root causes} (data and
configuration choices).

\smallskip\noindent
\textbf{DG-2: Invest in pre-evaluation stages.}
VA tools should provide richer support for data preparation and
learning configuration. Specifically, systems should visualize
data-quality indicators (missingness, class balance, label agreement,
feature correlations) \emph{before} model training begins, and
support interactive data cleaning, augmentation, and partitioning as
first-class operations.

\paragraph{Action types are dominated by hyperparameter adjustment.}
Among the 118~development-steering workflows, \emph{Adjust
Hyperparameters} is by far the most common action (39~workflows),
followed by \emph{Adjust Features}~(28), \emph{Select Samples}~(25),
and \emph{Edit Data}~(20). More structurally impactful actions---
\texttt{MT\_ModifyArch} (${\approx}$15~papers),
\texttt{MT\_ModifyObjective} (${\approx}$10~papers), and
\texttt{MT\_SelectAlgorithm} (${\approx}$12~papers)---are comparatively
rare. This indicates that VA systems are better at supporting
\emph{fine-tuning within a fixed design} than \emph{fundamental design
decisions}.

\smallskip\noindent
\textbf{DG-3: Support structural model changes, not only parameter tuning.}
Future systems should support higher-impact actions: architecture
modification (e.g., visual layer editors), algorithm switching (e.g.,
side-by-side comparison of model families), and objective function
editing (e.g., interactive loss-landscape exploration). These correspond
to rarely observed action types in our coding scheme
(\texttt{ACT\_TYP}~codes 7, 9, and~10).

\paragraph{Iteration history is rarely tracked.}
Only ${\approx}$15~papers in the corpus use \texttt{Run~Manage}
(\texttt{INT\_TYP\_MLDv}, code~3). Yet ML development is inherently
iterative: developers try different configurations, compare runs, and
sometimes revert to previous states.

\smallskip\noindent
\textbf{DG-4: Build iteration provenance into the system.}
VA systems should track and visualize the \emph{history} of steering
decisions, showing how successive human interventions changed model
behavior. This enables users to compare runs, identify productive
vs.\ counterproductive changes, and learn from their own
experimentation patterns.

\subsection*{C.2\quad Pathway-Specific Implications}

\paragraph{(b) Development Workflow Steering ($n{=}118$).}
Beyond the cross-pathway issues above, a notable finding is that
common statistical charts dominate visualization design (69\% of all
papers), even for complex model families. While scatter plots and bar
charts are versatile, they may not be the most effective
representations for understanding, e.g., training dynamics of deep
networks or the structure of ensemble models. In contrast, process/flow
visualizations, which could convey the iterative nature of ML
workflows, remain underexplored (Fig.~2(c)).

\smallskip\noindent
\textbf{RC-1:} What visual representations are most effective for
supporting architecture-level decisions (e.g., adding layers, changing
activation functions)? The current concentration on generic chart types
may be insufficient for structural model understanding.

\smallskip\noindent
\textbf{RC-2:} How can visualization effectively communicate the
\emph{expected downstream impact} of an upstream decision (e.g.,
removing 5\% of training data) on model performance, \emph{before}
retraining?

\paragraph{(c) Label Annotation ($n{=}36$).}
Among the 36~systems supporting label-related interaction, 22~do not
directly visualize either ground-truth or predicted labels. Annotators
in these systems work without context about what the model currently
``thinks.'' Furthermore, active learning (\texttt{INT\_TYP\_AnF},
code~2) appears in only ${\approx}$8~papers, meaning that most
annotation systems present data in a passive, user-driven sequence
rather than intelligently prioritizing which instances most need human
labels.

\smallskip\noindent
\textbf{DG-5: Always show prediction context during annotation.}
Even a simple overlay of the model's current prediction with its
confidence alongside the annotation interface would help annotators
focus on disagreements and uncertainties.

\smallskip\noindent
\textbf{DG-6: Integrate active learning as a default interaction
pattern.}
Rather than letting users browse randomly, VA annotation systems should
recommend which instances to label next, based on model uncertainty,
expected information gain, or diversity sampling.

\smallskip\noindent
\textbf{RC-3:} What is the optimal balance between showing model
predictions (which may anchor or bias annotators) and hiding them
(which may waste effort on ``easy'' instances)?

\smallskip\noindent
\textbf{RC-4:} How can visualization support annotation quality
control---detecting annotator fatigue, inconsistency, or systematic
bias---in real time?

\paragraph{(d) NLP Tasks ($n{=}40$).}
This pathway reveals two distinct interaction modes: development-oriented
(37.5\%) and deployment-oriented (30\%), with the latter featuring
language prompts (\texttt{INT\_MOD}~=~3) and prompt editing
(\texttt{KE\_EditPrompt}) as primary mechanisms. However, prompt
engineering remains visually unsupported in most systems: users are
given a text box and an output display, but there is almost no
visualization of \emph{why} a prompt worked or failed---no attention
visualization tied to prompt tokens, no comparison of prompt variants,
no visualization of the prompt's effect on the model's internal
representations.

\smallskip\noindent
\textbf{DG-7: Visualize prompt--output causality.}
For prompt-based systems, show which parts of the prompt influenced
which parts of the output (e.g., via attention attribution,
counterfactual analysis, or gradient-based saliency). This turns prompt
editing from trial-and-error into informed refinement.

\smallskip\noindent
\textbf{DG-8: Bridge the development--deployment gap.}
No system in the corpus connects deployment-mode NLP (using a model via
prompts) and development-mode NLP (training a model) within a single
workflow. Yet in practice, prompt-engineering insights often feed back
into fine-tuning decisions. Systems should support translating prompt
failures into fine-tuning directives within the same interface.

\smallskip\noindent
\textbf{RC-5:} How should prompt editing be visually supported when
prompts become long and structured (e.g., system prompts, few-shot
chains)? What are the visual analogues to code IDEs for prompt
engineering?

\smallskip\noindent
\textbf{RC-6:} Can visualization help users understand the
\emph{prompt search space}---i.e., which dimensions of variation in a
prompt matter most for output quality?

\paragraph{(e) Actions Outside VA ($n{=}23$).}
In this pathway, VA facilitates insight generation but actual model
modifications happen externally---typically via programming, configuration
files, or notebooks. The dominant forms of external knowledge injection
are parameter setting (13~workflows), code specification~(11), and
direct data input~(11). This decoupling between perception and action
represents a fundamental limitation: VA can show users \emph{what is
wrong}, but if the needed action is ``rewrite the loss function'' or
``restructure the data pipeline,'' the system offers no pathway to
execute it.

\smallskip\noindent
\textbf{DG-9: Embed lightweight code editors within VA systems.}
For actions that require code (e.g., modifying a loss function, editing
a data-transformation pipeline), provide an integrated notebook-style
interface. The key is not to replace full IDEs but to support
\emph{targeted edits motivated by visualization insights} without
breaking the analytical flow.

\smallskip\noindent
\textbf{DG-10: Design ``action templates'' for common external
modifications.}
Instead of requiring users to write code from scratch, offer
parameterized action templates (e.g., ``change learning-rate schedule
to $X$,'' ``add dropout layer with rate $Y$'') that can be triggered
from the VA interface and translated into the appropriate
code/configuration changes.

\smallskip\noindent
\textbf{RC-7:} What is the right level of abstraction for code-level
model modifications within a VA interface? Too low (raw code) negates
the purpose of VA; too high (simple sliders) lacks expressiveness.

\paragraph{(f) Explanation-Only Systems ($n{=}28$).}
These systems help users understand model behavior but provide no
mechanism to \emph{act on that understanding}. The paper describes
this as ``deferred knowledge injection,'' but the risk is that
knowledge is simply lost if no action channel exists. Among the
analytical patterns (\texttt{VIS\_PTN}) observed in this pathway,
\emph{Model Behaviour} and \emph{Error/Failure} dominate, while more
inherently actionable patterns---such as \emph{Trade-Off},
\emph{Drift/Shift}, or \emph{Bias/Fairness}---which
\emph{demand} a response, are less common.

\smallskip\noindent
\textbf{DG-11: Every explanation should suggest an action.}
At minimum, explanation-only systems should include ``next step''
recommendations: ``This cluster of errors is associated with feature $X$
$\rightarrow$ consider removing/transforming feature $X$,'' or
``Performance degrades for subgroup $Y$ $\rightarrow$ consider
collecting more $Y$ data.'' Even when the action is not automated,
making it explicit bridges the insight--action gap.

\smallskip\noindent
\textbf{DG-12: Shift from ``explain the model'' to ``explain what to
change.''}
Instead of showing attention maps or feature-importance scores (which
are model-state descriptors), design explanations that surface
\emph{actionable diagnostic patterns}: which training examples are most
harmful ($\rightarrow$ remove/relabel), which features are redundant
($\rightarrow$ prune), which hyperparameters are in a poor regime
($\rightarrow$ adjust).

\smallskip\noindent
\textbf{RC-8:} How do we evaluate whether an explanation-only system
actually improves downstream ML outcomes, given that actions happen
outside the system and may be delayed by hours or days?

\smallskip\noindent
\textbf{RC-9:} Is the explanation-only pattern sometimes
\emph{correct}---i.e., are there situations where understanding without
action is the appropriate design choice (e.g., regulatory auditing,
third-party monitoring)?

\subsection*{C.3\quad Summary}

Table~\ref{tab:implications} provides a compact overview of all
design guidelines and research challenges. Together, they aim to
transform the pathways identified in Section~5 from retrospective
descriptions of \emph{what has been built} into prospective guidance
for \emph{what should be built next}.

\begin{table}[ht]
\centering
\caption{Summary of design guidelines (DG) and research challenges (RC)
derived from the knowledge injection pathways.}
\label{tab:implications}
\footnotesize
\renewcommand{\arraystretch}{1.15}
\begin{tabular}{@{}p{0.62\columnwidth}p{0.30\columnwidth}@{}}
\toprule
\textbf{Guideline / Challenge} & \textbf{Source Pathway} \\
\midrule
DG-1: Design for action from the outset & Cross-pathway \\
DG-2: Invest in pre-evaluation stages & Cross-pathway \\
DG-3: Support structural model changes & Cross-pathway \\
DG-4: Build iteration provenance & Cross-pathway \\
DG-5: Show prediction context in annotation & (c) Annotation \\
DG-6: Integrate active learning by default & (c) Annotation \\
DG-7: Visualize prompt--output causality & (d) NLP \\
DG-8: Bridge development--deployment gap & (d) NLP \\
DG-9: Embed lightweight code editors & (e) Outside VA \\
DG-10: Offer parameterized action templates & (e) Outside VA \\
DG-11: Suggest actions from explanations & (f) Explanation \\
DG-12: Explain what to change, not only why & (f) Explanation \\
\midrule
RC-1: Vis.\ for architecture-level decisions & (b) Dev.\ steering \\
RC-2: Previewing impact of upstream changes & (b) Dev.\ steering \\
RC-3: Prediction context vs.\ annotator bias & (c) Annotation \\
RC-4: Real-time annotation quality control & (c) Annotation \\
RC-5: Visual IDEs for prompt engineering & (d) NLP \\
RC-6: Visualizing the prompt search space & (d) NLP \\
RC-7: Abstraction level for in-VA code edits & (e) Outside VA \\
RC-8: Evaluating explanation-only systems & (f) Explanation \\
RC-9: When is no-action the right design? & (f) Explanation \\
\bottomrule
\end{tabular}
\end{table}

\newpage

\end{document}